\documentclass{article}

\usepackage{subfig}

\usepackage{amsmath}
\usepackage{amsfonts}
\usepackage{amssymb}

\usepackage[final]{changes}
\usepackage[square,sort,comma,numbers]{natbib}

\newcommand{\bs}{\boldsymbol}
\newcommand{\bfu}{\boldsymbol{u}}
\newcommand{\bfr}{\boldsymbol{r}}

\newcommand{\bfb}{\boldsymbol{b}}

\newcommand{\vol}{V}

\usepackage{hyperref}

\title{ A unified derivation of Voronoi, power, and finite-element Lagrangian computational fluid dynamics }

\author{Daniel Duque \\
	CEHINAV Research Group,\\
	ETS Ingenieros Navales, Universidad Polit\'ecnica de Madrid,\\
	28040 Madrid, Spain
}

\begin{document}
	\maketitle

\begin{abstract}

Most approaches in Lagrangian fluid dynamics simulations proceed from the definition of particle volumes, and derive discrete versions of the spatial differential operators from it.

Recently, Gallouët and Mérigot \cite{gallouet_lagrangian_2018}  tackled simultaneously physical dynamics  and geometrical optimization, with the result that the pressure field is linked to a geometric feature: the weights of a power diagram. Their resulting dynamics, surprisingly, does not feature a pressure gradient, but spring-like forces between each particle and the centroid of its cell.

Inspired by this work, both geometrical and mechanical optimization are here included within a framework due to Arroyo and Ortiz \cite{arroyo_local_2006}. In a systematic way, we first find a connection with the smoothed particle hydrodynamics method. In what we will call the ``low-temperature limit'', we show that the requirement of zeroth order consistency leads to the Voronoi diagram, and a pressure field enforcing incompressibility leads to Gallouët and Mérigot's method.

If the requirement of first order consistency is added, the particle finite element method (pFEM) is recovered. However, it features an additional spring-like term that has been missing from previous formulations of the method.

Different methods are tested on two standard inviscid single-phase cases:the rotating Gresho vortex \added{and the Taylor-Green vortex sheet}, showing the superiority of pFEM, which is slightly increased by the additional force found here.

\end{abstract}


\section{Introduction}

Space must be discretized when the equations of fluid dynamics are simulated numerically. Computational fluid dynamics (CFD) may be broadly split in two classes. In Eulerian simulations, the discretization is fixed, and a careful spatial partition is usually devised before the simulation is run. However, in \added{other sort of} simulations the partition changes with time. If the shape of the partition is defined by points that move following the velocity field, one has a Lagrangian simulation, and the point are called particles, given their similitude with the infinitesimal particles used in continuum theory. The defining equation of motion for a Lagrangian simulation therefore states that the particles' position must change according to the velocity field:
\begin{equation}\label{eq:Lagrangian_u}
	\frac{d \bs{r}_i}{dt} = \bfu_i .
\end{equation}

From the definition of the volume of a particle, the spatial differential operators that appear in the equations of motion can be derived or, at least, guessed. For instance, the rate of change of volume of a particle may be written as
\begin{equation}\label{eq:discr_continuity}
	\frac{d \vol_i }{dt}  =
	\sum_j
	\frac{\partial \vol_i }{\partial \bfr_j} \cdot \frac{d \bfr_j}{dt} =
	\sum_j
	\frac{\partial \vol_i }{\partial \bfr_j} \cdot \bfu_j ,
\end{equation}
in which we acknowledge that the volume $V_j$ depends on the position of the particles (in a manner that depends on the particular approach), and we have used the Lagrangian condition of Eq.~\eqref{eq:Lagrangian_u}.

A comparison with the expression for the rate of dilation of a particle (which is used e.g. when deriving the continuity equation \cite{white_viscous_2006} ),
\begin{equation}\label{eq:incomp_V}
	\frac{d V}{dt} = V \,  \operatorname{div} \bfu ,
\end{equation}
leads to a discrete version of the divergence operator:
\begin{equation}\label{eq:discr_div}
	(\operatorname{div} \bfu)_i =
	\frac1{V_i}
	\sum_j \frac{\partial V_i }{\partial \bfr_j} \cdot \bfu_j .
\end{equation}
We will be considering the incompressible case here, in which volumes should not change. This means the latter expression should be brought to zero, or else the volumes must be kept fixed by some other means.

The gradient operator can be found by recalling that Euler's momentum equation for inviscid fluids,
\begin{equation}\label{eq:Euler}
	\rho \frac{d \bfu}{dt}  =  -\operatorname{grad} p ,
\end{equation}
can be obtained by Newton's Second Law applied to a particle \cite{white_viscous_2006}. This, in turn, can be derived from the Lagrangian, which involves the kinetic energy $T$ and the pressure energy $U_p$ (a Hamiltonian approach is of course equally valid, but less convenient since constraints will be later added). The pressure energy term is due to compression of the particles: $U_p = -\sum_j p_j V_j$, where $p_j$ is the pressure of particle $j$, and $V_j$ its volume. The Lagrangian will therefore be
\[
L = T - U_p = \frac12 \sum_j m_j u_j^2 + \sum_j p_j V_j .
\]
By the Euler-Lagrange equation, the change in velocity is given by
\begin{equation}\label{eq:discr_Euler}
	m_i \frac{d \bfu_i}{dt}  =  \sum_j \frac{\partial V_j}{\partial \bfr_i}  p_j.
\end{equation}
Comparing this against Euler's equation~\eqref{eq:Euler}, it is clear that the prescription for the discrete gradient is
\begin{equation}\label{eq:discr_grad}
	(\operatorname{grad} p)_i = - \frac1{V_i}
	\sum_j \frac{\partial V_j }{\partial \bfr_i} p_j .
\end{equation}

The gradient and the divergence will be the main operators we will discuss here,
\added{even though the role of the Laplacian in the determination of the pressure field will be considered --- it would of course also be present in the viscosity term, but this is not considered here}. Notice that the \deleted{the} pair of operators \eqref{eq:discr_div} and \eqref{eq:discr_grad} automatically satisfy conservation
of mechanical energy (within the accuracy of the time integrator) \cite{cercos-pita_role_2022}.

A convenient way to categorize Lagrangian methods is therefore by the definition of $V_i$, the volume associated with particle $i$. For instance, in Smoothed Particle Dynamics (SPH), a particle is associated with a kernel function $W_i(\bfr) = W(|\bfr - \bfr_i |)$, which has a bell shape with some typical range, larger than the mean spacing between particles. The method is usually derived by defining SPH interpolation as $\bar{A} (\bfr) = \sum_j A_j W_j(|\bfr - \bfr_j |) V_j$, where $A_j$ is the value of any field at particle $j$, and $V_j$ its volume. The latter may be defined, in a rather circular fashion, by considering $A_j = \rho_j$, the density of particle $j$. Then,
$\bar{ \rho} (|\bfr|) = \sum_j m_j W(|\bfr - \bfr_j |) $ . Evaluating the former interpolation function at $\bfr = \bfr_i$, we get $\bar{\rho}_i = \sum_j m_j W(|\bfr_i - \bfr_j |) $, thus defining the density of particle $i$ as a function of how ``crowded'' it is by other particles. The volume of particle $i$ is finally $V_i= m_i / \bar{\rho}_i$.

This results  
in a very simple and versatile method with many advantages \cite{violeau_fluid_2012,monaghan_smoothed_2012}. For instance, the resulting forces look similar to the usual inter-particle forces of molecular dynamics, and satisfy action and reaction.
Also, the method is completely mesh-less, with no connectivity information to keep track of. Unfortunately, it also lacks some basic requirements. For example, the sum of all the volumes is not, in general, equal to the total volume. This is related to the fact that the method  lacks ``zero-th order consistency'': if a constant function is interpolated as $\bar{A}(|\bfr|) = A \sum_j  W_j(|\bfr|) V_j$, the result is not constant (although it does tend to a constant as the number of particles increases as the range of the kernel is fixed.)
The method is often kept simple by allowing a small compressibility, in the family called Weakly Compressible SPH --- even though incompressible SPH is also a possibility \cite{cummins_sph_1999,ellero_incompressible_2007}.

Another natural method to discretize space is to take each of the particles as the seeds of a Voronoi diagram \cite{serrano_voronoi_2005}: space is partitioned into Voronoi cells --- cell $i$ being defined as the spatial region with points closer to particle $i$ than to any other. The result does comply with zero-th order consistency, and the volume is exactly the sum of all cells' volume. It is not difficult to set up a simulation where the particles' velocities evolve following the gradient of the pressure field, as in Eq.~\eqref{eq:discr_Euler}. The latter may be fixed by requiring incompressibility: the vanishing of Eq.~\eqref{eq:discr_continuity}. This may be achieved by a Newton-Raphson method in which a Laplacian appears as the divergence of the gradient in a pressure Poisson equation (PPE) --- a standard technique in incompressible CFD:
\begin{equation}\label{eq:PPE}
	(\operatorname{div} \operatorname{grad}  p )_i =
 - \frac{\rho}{\Delta t} (\operatorname{div}  \bfu)_i ,
\end{equation}
with appropriate expressions for the differential operators as in Eqs.~\eqref{eq:discr_div} and \eqref{eq:discr_grad}, which involve geometric features of the diagram.
This method will be called VPD (Voronoi Particle Dynamics) and will be tested in Section \ref{ss:simulations:VPD}.

A related approach is to consider the dual of the Voronoi diagram, the Delaunay triangulation. The volume of particle $i$ can be taken as the volume of a prism of unit height on $i$, with the base formed by all the triangles incident to $i$. This is the usual finite element method (FEM) C1 shape function. The resulting method is basically the pFEM method. Notice that in 2D the shape functions of particles joined by a triangulation edge overlap on the two triangles shared by them (the analogous shape functions for the Voronoi diagram would be just plateu functions, with a value of $1$ in the corresponding cell, and $0$ outside, and no overlap.) This method, and its variants, has been extensively used (see e.g. \cite{cremonesi_state_2020} for a review), with good results. Incompressibility is usually taken care of through a PPE identical to Eq.~\eqref{eq:PPE}, of course with the appropriate definitions of the divergence and gradient operators.
This pFEM method will be deduced within a general framework in Section \ref{ss:pFEM}, and tested in Section \ref{ss:simulations:pFEM}.

Recently, de Goes \text{et al.} \cite{de_goes_blue_2012,de_goes_power_2015} proposed an alternative partition of space based on power diagrams (also called Laguerre diagrams, or Dirichlet cell complexes). These are similar to Voronoi diagrams, but each particle has an additional quantity, its weight, that can change its volume from the one it would have in the plain Voronoi diagram. Within this approach, incompressibility can be enforced by adjusting the weights in order the volumes in the diagram be equal. (For the latter, as we will discuss
in Section \ref{ss:power_particles}, another differential operator is needed, which is a sort of discrete Laplacian.)
It is shown that a certain ``geometrical'' energy is minimized when the volumes are equalized and the diagram is centroidal \cite{du_centroidal_1999} (i.e. the position of the particles coincide with the centroids of their cells.) This motivates an algorithm in which particles are placed at the centroid of their cells at each iteration. The pressure field is still computed by the same techniques as in VPD, and its gradient used in the equation of motion.
We will refer to this method as dG, and will test it in Section \ref{ss:simulations:DG}.

The fact that the pressure field is treated in dG as if enforcing incompressibility, when this requirement is actually achieved through the weights, is clearly odd, as is the placement of the particles at the centroids. This is due to the geometrical energy being treated separately from the physical energy. A possible solution was proposed by Gallouët and Mérigot \cite{gallouet_lagrangian_2018}, by treating both terms
simultaneously, as two contributions to a global energy. They show that the forceful placement of the particles at the centroids is replaced by a spring-like force between particles and centroids. Moreover, the pressure is seen to be basically equivalent to the weights.  This connection, surprisingly, forfeits the need for the gradient and divergence operators: the pressure gradient disappears from the equation of motion.
The resulting simulations are fast, simple, and versatile \cite{levy_fast_2020, levy_partial_2021}.
This method will be called GM, derived in Section \ref{ss:power_particles}, and tested in Section \ref{ss:simulations:GM}.


In retrospect, all approaches but GM proceed from a geometrical proposal for the particle volumes, and derive their dynamics from it. The latter authors tackle physical dynamics and geometrical optimization at the same time, with the result that a physical field, the pressure, is linked to a geometric feature, the weights.
Inspired by this work, we describe an approach that includes both geometrical and mechanical optimization. In this line of thought, the work by Arroyo and Ortiz  \citep{arroyo_local_2006} is particularly insightful in providing a link between geometrical constructions and optimization. We will make extensive use of their formalism, even if our results are restricted in this work to what we will call the ``low-temperature limit.'' It is likely that interesting methods can be formulated without reaching that limit, but we leave that work for the future.

We begin with a brief introduction to the formalism in Section~\ref{ss:entropy}, in which a connection with the SPH method is pointed out. The requirement of zeroth order consistency is seen to lead to the Voronoi diagram in Section~\ref{ss:Voronoi}. When a pressure field is introduced in Section~\ref{ss:power_particles}, Gallouët and Mérigot's method is recovered, as the only possible Voronoi-based method (at least within this framework).

If the requirement of first order consistency is further added,  Section~\ref{ss:Delaunay} the Delaunay triangulation is obtained, as is known from the work of Rajan \cite{rajan_optimality_1994}. If a pressure field is added, Section~\ref{ss:pFEM} we recover the pFEM method, but with an additional spring-like term that has been missing from previous formulations of the method.

\section{General framework}
\label{ss:entropy}

Following \citep{arroyo_local_2006}, let us consider a set of $N$ particles with positions $\{ \bfr_i\}$, each one endowed with a shape function $\phi_i(\bfr)$, which may be different for each particle and is, as yet, \emph{unknown}. The volume of a particle is given by the spatial integral of its shape function:
\begin{equation}\label{eq:V_as_int}
	V_i := \int  \phi_i(\bfr)   d \bfr  .
\end{equation}
Let us define a free energy
\begin{equation}\label{eq:cost}
	F_0 = \Pi \int \left[ \epsilon(\bfr) - s(\bfr) \right] d \bfr,
\end{equation}
where $\Pi$ is a constant with dimensions of pressure.
An important ingredient is the (dimensionless) energy function
\[
\epsilon(\bfr)= \beta \sum_i \phi_i(\bfr) \left| \bfr - \bfr_i  \right|^2 ,
\]
($\beta$ has units of inverse length squared).
Disregarding $s$, function $F_0$ is a sort of cost function that is found in many areas: e.g. in transport theory it represents the cost of bringing some abstract material from each point
in space to a given set of nodes \cite{peyre_computational_2019}. The goal is usually to minimize it, much
like a physical energy. Alternatively, we may think that by minimizing $F_0$ we are minimizing the second moment of the shape functions (the volumes being their zeroeth moment, as in Eq.~\eqref{eq:V_as_int}).

The second contribution to $F_0$ is a (dimensionless) entropic term:
\[
s(\bfr) = - \sum_i \phi_i(\bfr)  \ln \left[ \phi_i(\bfr)  \right].
\]
\replaced{
The idea is that, at each point $\bfr$, the value of each shape function
$\phi_i(\bfr)$ is required to be non-negative (as implicit by the $\log$ function), and the sum of all is required to be $1$, as discussed next. These are the usual requirements for a set of probabilities, hence the usage of what is basically Shannon's entropy of information theory (but for unimportant factors). Following Jaynes' principle, by maximizing it ``unbiased inference'' \cite{jaynes_information_1957} is achieved.
Thanks to the energy function, a result different from the trivial	 $\phi_i(\bfr) = 1/N$ may be found: intuitively,
a particle $i$ that is very close to $\bfr$ will have a larger value of $\phi_i(\bfr)$ than distant particles. Additional constraints will also affect these values. By considering all points in space, each shape function $\phi_i(\bfr)$ may be found. These are not probability densities, since they are not normalized (in fact, their integrals define the volumes of the particles, Eq.~\eqref{eq:V_as_int}.)
}
{
This term also has connections with other areas, such as information theory, where it corresponds to Shannon's entropy (but for the $-1$ sign, introduced here for simplicity and not important). Following Jaynes' principle, the goal is to maximize it, in order to achieve what is called ``unbiased inference'' \cite{jaynes_information_1957}.
}

\subsection{Voronoi diagrams}
\label{ss:Voronoi}

As just mentioned, a basic constraint is to ask for the shape functions to reconstruct a constant function (i.e. zeroth order consistency):
\[
\sum_i \phi_i(\bfr) = 1 \qquad \text{at every }\bfr .
\]
This may be enforced by extremizing
\[
F_1 =  F_0 + \Pi \int
\alpha(\bfr) \left( \sum_i \phi_i(\bfr) - 1\right)   d \bfr ,
\]
with a Lagrangian multiplier $\alpha(\bfr)$. We find its extreme with respect to $\phi_i(\bfr)$ when
\[
\phi_i (\bfr) = \exp\left( f_i(\bfr)  - \alpha(\bfr) -1\right) ,
\]
where we define
\[
f_i (\bfr) :=  - \beta \left| \bfr - \bfr_i  \right|^2 .
\]
This is the common Boltzmann probability if $f_i$ is identified as a dimensionless negative energy. 
The condition $\sum \phi_i(\bfr) = 1$ fixes $\alpha(\bfr) =  \ln Z(\bfr) - 1$, with the partition function
\[
Z(\bfr):=\sum_i \exp\left( f_i(\bfr) \right) ,
\]
and, finally,
\begin{equation}\label{eq:phi_general}
	\phi_i(\bfr) = \frac{1}{Z(\bfr)} \exp\left( f_i(\bfr) \right)  .
\end{equation}

\begin{figure}
	\centering
	\includegraphics[width=0.95\linewidth]{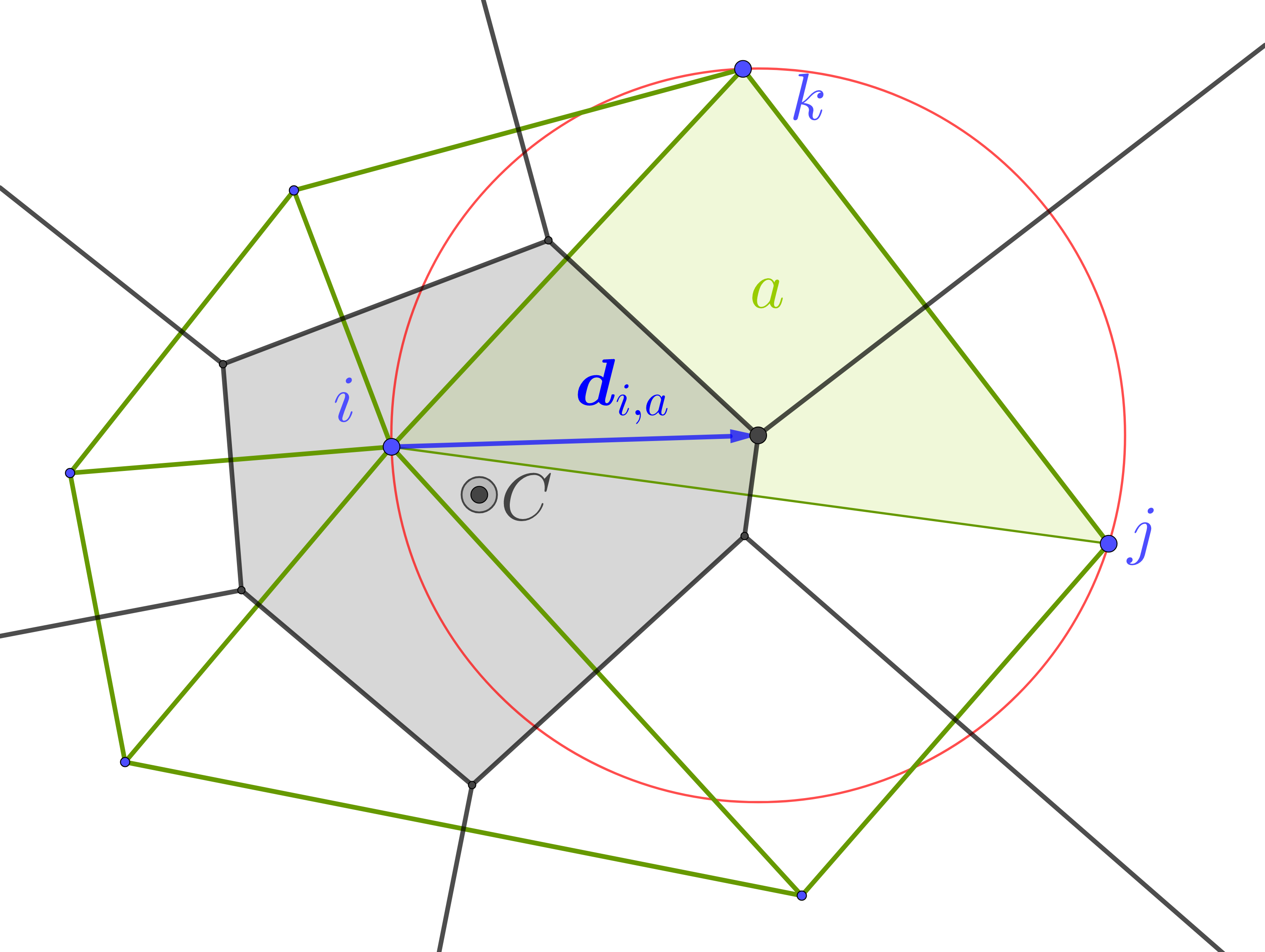}
	\caption{Sketch of the geometry of a typical particle configuration. Voronoi diagram in black, with cell corresponding to particle $i$ filled in gray. Point $C$ is the centroid of this cell. In green, the dual Delaunay triangulation, of which particles $i$ and $j$ share two triangles, where triangle $a$ is highlighted. The circumcenters of the triangles correspond to vertices of the Voronoi Diagram. The circumcircle of triangle $a$ is highlighted in red. Vector $\bs{d}_{i,a}$ goes from particle $i$ to the circumcenter of triangle $a$. }
	\label{fig:geometry}
\end{figure}

The Voronoi construction is recovered in the ``cold'' limit (large $\beta $): the functions $\exp\left( f_i(\bfr) \right)$ will become very thin Gaussians, with very little overlap, hence $Z(\bfr)$ becomes just $\exp\left( f_i(\bfr) \right)$ in the space domain closer to $i$ than to any other particle. This is precisely the definition of a Voronoi cell, with the ratio $\exp\left( f_i(\bfr) \right) / Z(\bfr) $ being equal to $1$ within the cell, and $0$ outside. This means our shape functions are flat plateaus of value $1$. In Fig.~\ref{fig:geometry} a sketch  of a typical configuration is provided. The cell of particle $i$ is highlighted as a gray region, and the shape function would be equal to $1$ inside that region, and $0$ outside.

Notice that, as pointed out in \cite{arroyo_local_2006}, in a ``warm'' situation an SPH formulation may be recovered. If $\beta \sim 1/h^2$, where $h$ is some smoothing length larger than the mean distance between particles, the shape functions become SPH kernels. More specifically, these kernels would be Gaussian, and would feature a Shepard correction to ensure zeroth order consistency. This connection has not been fully explored, to our knowledge.

Dynamics may be obtained from the Euler-Lagrange equations
\[
m_i \frac{d \bfu_i}{d t} = \frac{\partial L}{\partial \bfr_i} ,
\]
with the Lagrangian
\[
L =  \frac12 \sum_i m_i u_i^2  -F'_1  ,
\]
where $F'_1$ the free energy $F_1$ evaluated at the extremizing values of $\phi_i$:
\begin{equation}\label{eq:extremized_Lagrangian}
	F'_1 = \Pi \int   \ln \left[  Z (\bfr) \right] d \bfr .
\end{equation}

The force upon particle $i$ is given by
\begin{align*}
  \frac{\partial L}{\partial \bfr_i} &=
  -
\frac{\partial F'_1}{\partial \bfr_i} =
\Pi
\frac{\partial }{\partial \bfr_i}
\int \ln \left[  Z (\bfr) \right] d \bfr
= \\
& =
\Pi
\int  \frac{1}{Z} \frac{\partial Z(\bfr) }{\partial \bfr_i} d \bfr =
\Pi
\int \sum_j \phi_j(\bfr) \frac{\partial f_j(\bfr) }{\partial \bfr_i} d \bfr,
\end{align*}
where we have used the fact that $\bfr_i$ only appears in the $\exp(f_i)$ term in $Z$.
Hence, quite generally
\begin{equation}\label{eq:quite_general_eom}
	m_i \frac{d \bfu_i}{d t} = \Pi
	\int \sum_j \phi_j(\bfr) \frac{\partial f_j (\bfr)}{\partial \bfr_i} d \bfr.
\end{equation}

In this case, the only $f_j(\bfr)$ that depends on $\bfr_i$ is $f_i(\bfr)$, hence
\[
\frac{\partial L}{\partial \bfr_i} =
2\Pi \beta
\int \phi_i(\bfr) [\bfr - \bfr_i ] d \bfr =
2\Pi \beta   V_i (\bfb_i - \bfr_i ) ,
\]
where we have defined the centroid (or, center of mass) of particle $i$'s cell as the normalized first moment of its shape function:
\[
 V_i \bfb_i := \int \phi_i(\bfr) \bfr \, d\bfr .
\]

Our final equation of motion is, therefore
\[
m_i \frac{d \bfu_i }{dt } =  2\Pi \beta V_i (\bfb_i - \bfr_i ) ,
\]
or, given that $\rho$ is constant, and $m_i = \rho V_i$
\begin{equation}\label{eq:spring_to_centroid}
	\frac{d \bfu_i }{dt } =  \omega^2 (\bfb_i - \bfr_i ) ,
\end{equation}
a spring connecting each particle to the center of mass defined by its shape function. This make sense, since it is well known that the lowest energy is achieved in a diagram where the position of each node coincides the centroid of its Voronoi cell: a centroidal Voronoi diagram. The resulting motion consist of vibrations around these minima. The centroid of cell $i$ is shown in Fig.~\ref{fig:geometry} as point $C$, and particle $i$ would be radially attracted towards it by the spring force.

The spring angular frequency, $\omega$ , equals $ \sqrt{ 2\Pi \beta / \rho}$. We therefore have the freedom of uncoupling the value of $\beta$ (how localized our shape functions are) and the spring frequency. This holds even in the very large $\beta$ limit, which is the one targeted: even if this quantity is \replaced{large}{great}, we can still choose a small value of $\Pi$, so that $\omega$ has any value we want. The resulting oscillatory motion would not really correspond to a physical system, since the pressure has not been added --- this will be introduced in the next section. \added{If some dissipation is added, the resulting motion could be used to construct centroidal diagrams, as a smoother version of Lloyd's algorithm.}

\subsection{Power particles from optimization}
\label{ss:power_particles}

The pressure may be introduced through an incompressibility constraint in the Lagrangian:

\[
L =  \frac12 \sum_i m_i u_i^2  -F_1 +   \sum_i p_i (V_i - V_i^0)
\]

By writing
\(
\sum_i p_i V_i = \sum_i p_i \int \phi_i (\bfr) d\bfr ,
\)
we can write the equivalent
\[
L =  \frac12 \sum_i m_i u_i^2  - F_2 , 
\]
where
\[
F_2 := F_1 - \int \sum_i p_i  \phi_i(\bfr) d\bfr +  \sum_i p_i V_i^0 .
\]

Extremization with respect to the values of the shape functions leads to
the same expression as in Eq. \eqref{eq:phi_general}, but now
\[
f_i(\bfr) =  - \beta \left| \bfr - \bfr_i  \right|^2  + \frac{ p_i }{ \Pi  }  .
\]
In the large $\beta$ limit, the range of influence of particle $i$ will be region of points with smaller value of $\left| \bfr - \bfr_i  \right|^2  - p_i / (\Pi \beta)  $. This is precisely  the definition of a power diagram, with weights $w_i = p_i / (\Pi \beta) $. This means
\begin{equation}\label{eq:GM_pressure_weight}
	p_i = \frac12 \rho \omega^2  \, w_i ,
\end{equation}
exactly Gallouët and Mérigot's (GM) result.
The pressure is therefore coupled with the weights, and the latter may be obtained by enforcing incompressibility by a volume equalizing procedure. This may be carried out with a Newton-Raphson procedure:
\begin{equation}\label{eq:NR}
	\frac{\partial V_i}{\partial w_j} \delta w_j = V_i^0-V_i ,
\end{equation}
where the discrete differential operator is closely related to a well-know expression for the Laplacian in the FEM, as discussed in \cite{de_goes_blue_2012, de_goes_power_2015, gallouet_lagrangian_2018}:
\begin{equation}\label{eq:laplacian}
	\Delta_{ij} = - \frac{1}{2} \frac{\partial V_i}{\partial w_j}.
\end{equation}
Notice that the resulting dynamics are still as in Eq. \eqref{eq:spring_to_centroid}: the pressure does not appear explicitly in the equations of motion. The difference with that section is the requirement of volume equalization in the GM method.

\added{Notice the resulting dynamics will not conserve kinetic energy, since this will be used in pushing and pulling the springs. It is not clear if the latter movement can be cast as a conservative force, so a general mechanical energy may not be conserved. Neither linear nor angular momentum can be expected to be conserved, since the spring forces are applied from the centroids, not between particles.}

\subsection{Delaunay triangulation and FEM shape functions}
\label{ss:Delaunay}

Let us temporarily forget incompressibility and go back to a geometric requirement:
that linear functions also be exactly reproduced (``first order consistency''): $\sum \bfr_i \phi_i(\bfr)  = \bfr $.
The free energy function would be,
\[
F_3 := F_1 -
\Pi \int
\boldsymbol{\lambda}(\bfr) \cdot \sum \phi_i\bfr_i \left[  \bfr - \bfr_i  \right]  d\bfr
\]
with $\boldsymbol{\lambda}$ being a Lagrangian multiplier vector that enforces
linear consistency (with units of inverse length).

Minimization with respect to $\phi_i$ produces the same expression as
Eq.~\eqref{eq:phi_general}, but with
\[
f_i (\bfr) =
- \beta \left|  \bfr - \bfr_i  \right|^2
+ \boldsymbol{\lambda}(\bfr) \cdot \left(  \bfr - \bfr_i  \right)
\]

At ``low temperatures'', the $\boldsymbol{\lambda}$ function will try to bring the $\beta \left|  \bfr - \bfr_i  \right|^2$ term close to some point that is close to the particles, in some sense. In 1D, this is the mid-point between neighboring particles. In 2D, three nodes are involved. Since the $f_i$ should approximately be quadratic in this limit, of the form $\sim \beta(\bfr-\bfr_\text{c})^2$, the iso-contours of this function are circles around $\bfr_\text{c}$, and the three nodes should have the same value: i.e. they should lie in a circle centered in $\bfr_\text{c}$ --- precisely the definition of the circumcircle of a triangle. Moreover, no other particle should be within this circle, which precisely identifies the triangulation as the Delaunay triangulation. A rigorous proof of this fact was given by Rajan \cite{rajan_optimality_1994}.

Moreover, the combination of zeroth and first consistency dictates that the resulting shape functions are the well-known C1 FEM functions: piece-wise linear functions with a value of one at their corresponding node, going to zero at its neighbors. The latter are, in 1D, the particles at right and left, and in 2D the collection of particles belonging to the incident triangles in the Delaunay triangulation. The 3D case is similar, with the Delaunay tetrahedralization. We have excluded degenerate cases, in which more than three points lie in a circle and the Delaunay triangulation is not uniquely defined --- however, this procedure still provides well-defined shape functions in these cases.

Fig.~\ref{fig:geometry} shows the Delaunay triangles incident to particle $i$. Each triangle contributes with a skewed prism of height $1$ and $i$, linearly decreasing to $0$ at the other two triangle vertices. With all the other incident triangles, the resulting shape function is a prism with height $1$ at $i$, and base equal to the union of all the triangles.

The partially extremized Lagrangian will now be as in Eq.~\eqref{eq:extremized_Lagrangian}, and the equation of motion, as in Eq.~\eqref{eq:quite_general_eom}.
The main difference is that we now find
\[
\frac{d f_j }{d \bfr_i} =
\left[  2\beta (\bfr-\bfr_i ) - \boldsymbol{\lambda} \right] \delta_{i,j} +
\frac{\partial \boldsymbol{\lambda} }{\partial \bfr_i} \cdot (\bfr-\bfr_j ) ,
\]
where care is taken of the fact that $\boldsymbol{\lambda}$ actually depends on the positions.

The sum at the integrand of Eq.~\eqref{eq:quite_general_eom} is therefore
\begin{align*}
& \sum_j \phi_j
\left[  2\beta (\bfr-\bfr_i ) - \boldsymbol{\lambda} \right] \delta_{i,j} +
\frac{\partial \boldsymbol{\lambda} }{\partial \bfr_i}
  \cdot (\bfr-\bfr_j ) = \\
& \left[  2\beta (\bfr-\bfr_i ) - \boldsymbol{\lambda} \right] \phi_i +
\frac{\partial \boldsymbol{\lambda} }{\partial \bfr_i}
  \cdot \sum_j \phi_j (\bfr-\bfr_j ) ,
\end{align*}
and the last term vanishes. We are therefore left with
\begin{equation}\label{eq:2beta_lambda}
	\frac{d L}{d \bfr_i} =
	\Pi \int \phi_i(\bfr) \left[  2\beta (\bfr-\bfr_i ) - \boldsymbol{\lambda}(\bfr) \right] d\bfr.
\end{equation}
In the large $\beta$ limit 
the integrand tends to a simple form in each of the triangles incident to $i$. In fact, it becomes a constant vector, whose value is $2\beta$ times the distance from $i$ to the circumcenter of the triangle. This is a simple, remarkable result that is, in some sense, the equivalent of the force tethering each particle with their Voronoi centroid in the previous Section~\ref{ss:power_particles}.
Fig.~\ref{fig:geometry} shows this vector for a selected triangle $a$,
called $\bs{d}_{i,a}$. Notice the circumcenter coincides with a Voronoi vertex: this fact establishes
an intriguing connection with the spatial operators obtained in Voronoi particles \cite{serrano_voronoi_2005}. Indeed, the same vector appears there, but other factors (distance
between particles, length of the Voronoi edge, areas of triangles\ldots ) complicates a clear correspondence.

This constant vector can be pulled out of the integral, leaving a trivial integral of the shape function. 
%
%
The equation of motion is, then,
\[
m_i \frac{d \bfu_i }{dt } =
2 \Pi  \beta  \sum_{a \in T(i)}  V_{i,a} \bs{d}_{i,a}  ,
\]
where the sum $a$ is on triangles incident to particle $i$ (the collection of which is represented by $T(i)$), each with FEM volume $V_{i,a}$ (equal to $1/3$ of the area of the triangle).

Writing $ m_i = \rho V_i$ does not quite cancel out the volume factors at
each side of the equation as in Eq.~\eqref{eq:spring_to_centroid}, and we find
\[
\frac{d \bfu_i }{dt } =
\frac{\omega^2}{V_i}
\sum_{a \in T(i)} V_{i,a} \bs{d}_{i,a}  ,
\]
where $\omega^2 := 2\Pi\beta /\rho $. The force is different for each particle $i$, and even for each of the triangles $a$ incident to particle $i$, due
to the $ V_{i,a} / V_i $ factor.

%

\subsection{Standard and corrected pFEM}
\label{ss:pFEM}

The addition of pressure in the particles can be taken into account as when deriving power particles, Section \ref{ss:power_particles}. The Lagrangian would now be
\begin{equation*}
	L =  \frac12 \sum_i m_i u_i^2  -F_3 +   \sum_i p_i (V_i - V_i^0) ,
\end{equation*}
from which we find
\[
f_i :=
- \beta \left|  \bfr - \bfr_i  \right|^2
+ \boldsymbol{\lambda}(\bfr) \cdot \left(  \bfr - \bfr_i  \right)
+ \frac{p_i}{\Pi} .
\]
Despite the appearance of the pressure in $f_i$, results just derived in the previous Section regarding the shape function do not change at all, at variance with power particles. The vector field $\boldsymbol{\lambda}(\bfr)$, however, does change, resulting in a force that incorporates the pressure:
\[
\frac{d \bfu_i }{dt } =
\frac{\omega^2}{V_i}
\sum_{a \in T(i)} V_{i,a} \bs{d}_{i,a}    -
\frac1\rho (\operatorname{grad} p)_i ,
\]
where the sum is over triangles incident to $i$. The discrete gradient operator is exactly as in the standard pFEM method, and as given by Eq.~\eqref{eq:discr_grad}.
This is not surprising since, if the shape functions and volumes do not change with the introduction of pressure, one can start
with a fixed prescription for the volumes, and derive the spatial differential operators from it (as e.g. in SPH and VPD).

The pressure is thus unrelated to the geometry, at variance with the GM method. Indeed, the Delaunay triangulation of a power diagram does not change with the weights at all (as long as they do not induce flips). The pressure field can then be obtained from the incompressibility enforcement. This, again, can be accomplished by a standard PPE procedure \added{similar to Eq.~\eqref{eq:PPE}.
The precise PPE solved actually uses the FEM Laplacian of \ref{eq:laplacian} instead of the divergence of the gradient. This is not completely in line with our discussion, but results are seen to be much better with this prescription.}

Given we are free to choose the strength of the additional force (again because even if $\beta$ is large, $\beta\Pi$ does not need to be), we may set it equal to zero, thus recovering the standard pFEM. If a non-zero force is considered, we will call the resulting method ``corrected pFEM''.

\added{About conservation properties: standard pFEM does conserve kinetic energy, since the pressure term appears as a Lagrangian constraint which should be zero. The resulting inter-particle forces moreover satisfy action and reaction, thus ensuring momentum is also conserved. Angular momentum is not conserved in general, since the forces are not central (they are not parallel with the line connecting two particles). Unfortunately, we have been unable to translate these conservation properties to corrected pFEM, even though e.g. kinetic energy will shown to be reasonably conserved in the following Section.}

\section{Simulations}
\label{ss:simulations}

\subsection{Gresho vortex}

\subsubsection{Setup}
\label{ss:simulations:setup}

We have performed 2D simulations with $49\times49= 2401$ particles on a unit-length square. Additional particles are placed along the sides to provide a boundary --- these will not move and are given null velocity and pressure values. Particles are initially placed on a square lattice,
and their positions are slightly perturbed following the initial velocity field times a small time $\delta t = 10^{-2}$. They are then subject to a Lloyds procedure of 1000 steps, in order to produce a centroidal initial diagram, which drives them from the square lattice to an approximately hexagonal one (corresponding to an equilateral Delaunay triangulation). \deleted{See Fig. \ref{fig:snap0} for a typical initial particle distribution.}

A velocity field is then imposed upon the particles. We have used the Gresho vortex, which has a piece-wise continuous tangential velocity field that depends on the distance to the center and is often used when testing CFD schemes \cite{liska_comparison_2003}. It goes from null at the origin to a value of $1$ at $r_0 = 0.2$ in the ``core'' region, so that all particles inside $r < r_0$ core rotate at the same angular velocity. It then goes back linearly to zero from $r_0$ to $2r_0$, in the ``corona'' region, and is null for larger distances:
\[
\bfu_{\phi}(r) =
\begin{cases}
	\frac{r}{r_0}  			& 0      \leq r < r_0  \\
	2 - \frac{r}{r_0}		& r_0    \leq r < 2 r_0   \\
	0						&  2 r_0 \leq r  .
\end{cases}
\]

The pressure field is obtained by solving for $\nabla p = u^2 /r$. Shifting its value so that $p=0$ at $r \geq 2r_0$:
\[
p(r)  =
\begin{cases}
	- 4 \ln 2  +  \frac52 + \frac12 \left[ \left( \frac{r}{r_0} \right) ^2 - 1 \right]
	& 0      \leq r < r_0 , \\
	4 \ln (\frac{r}{2 r_0}) -
	4 \left[  \left( \frac{r}{r_0} \right)  -2 \right]  +
	\frac12 \left[ \left( \frac{r}{r_0}  \right)^2 -4  \right]
	& r_0    \leq r < 2 r_0, \\
	0   &  2 r_0 \leq r.
\end{cases}
\]
The limit values of the pressure field are $0$ and $-4\ln(2) + 2 \approx -0.77$.
The pressure field is \emph{never} set, but obtained as a result, and compared with the latter, exact, expression.

The revolution period is $T= 2\pi \times 0.2 \approx 1.2566$, simulations are carried out for three complete periods. The time-step is set to $\Delta t = 0.005$, corresponding to a Courant number Co=$ 1\times 0.005 / (1/50) = 0.25$. We have selected five simulation schemes, which we describe separately.

\subsubsection{Voronoi particles}
\label{ss:simulations:VPD}

We first test a VPD approach, as the one in Ref. \cite{serrano_voronoi_2005}, with the Voronoi volume providing the gradient and pressure operators, and a PPE to find the pressure field. \added{This method is energy-conserving, in the same way as pFEM is (the pressure appears as a Lagrange multiplier). However, the resulting forces do not satisfy action and reaction in general, nor are they directed in the inter-particle direction. Hence, neither linear on angular momentum is conserved.}

The time integration is solved by a mid-point implicit scheme: particles are moved according to the previous velocity field, for a time $\Delta t /2$. The pressure is then obtained from the PPE as in Eq.~\eqref{eq:PPE}, since the resulting velocity field will have a non-zero divergence. The gradient of the pressure is used to improve the velocity in a second iteration of the method. When the predicted mid positions change little, the velocity field is accepted and a whole step with time $\Delta t$ is made.

\begin{figure}
	\centering
	\includegraphics[
	width=0.45\linewidth]{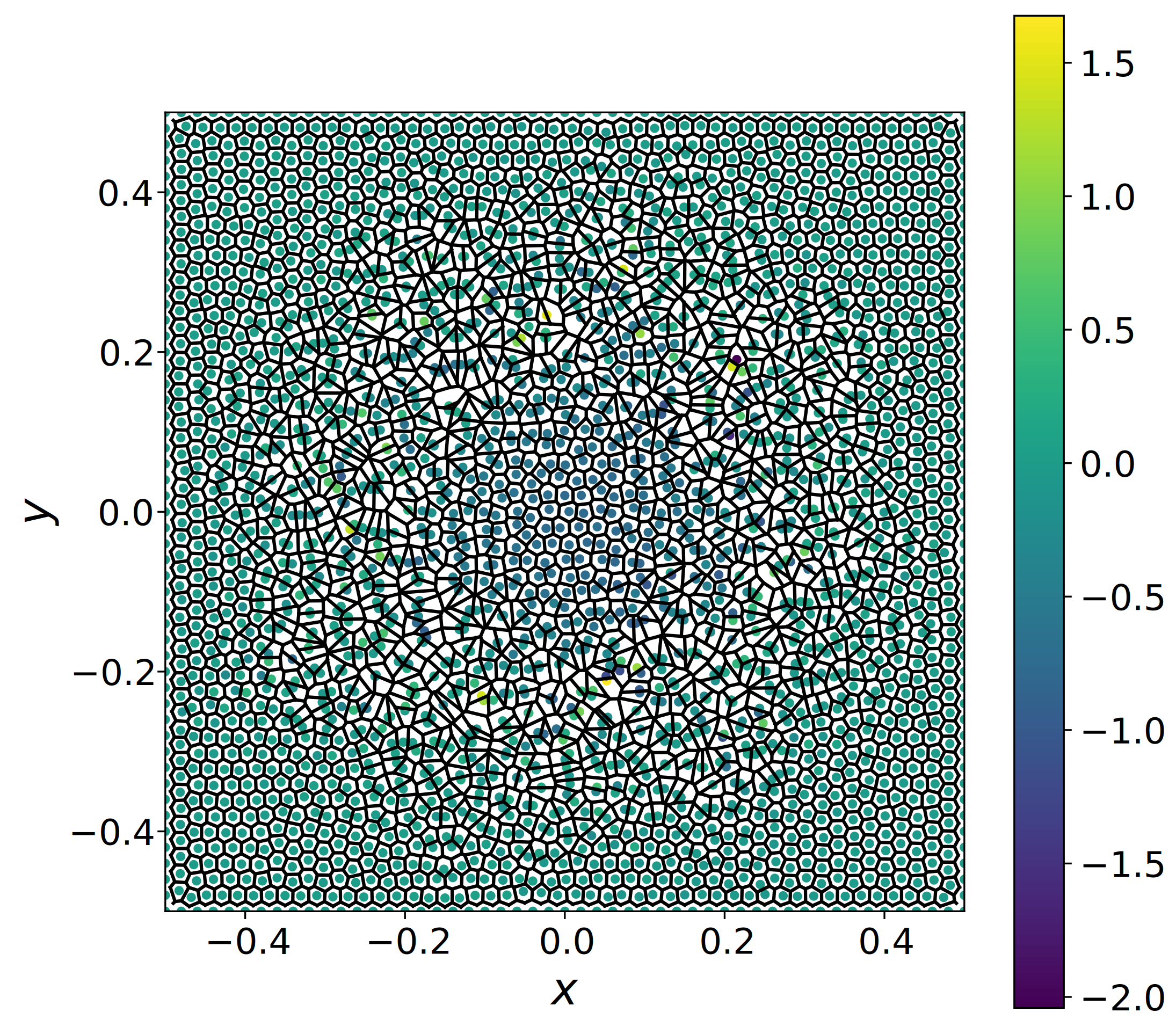} \hfil
	\includegraphics[
	width=0.45\linewidth]{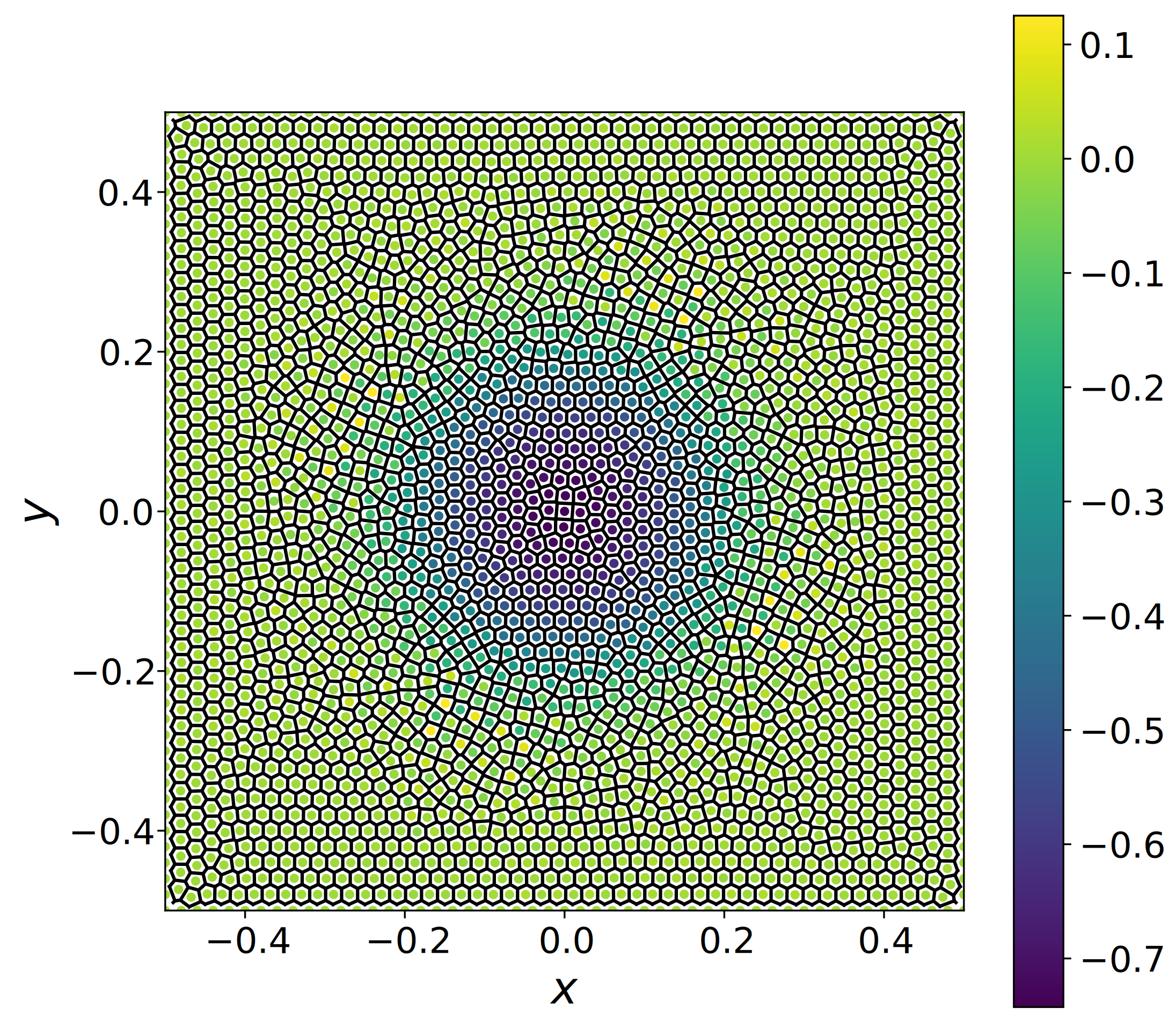} \\
	\includegraphics[
	width=0.45\linewidth]{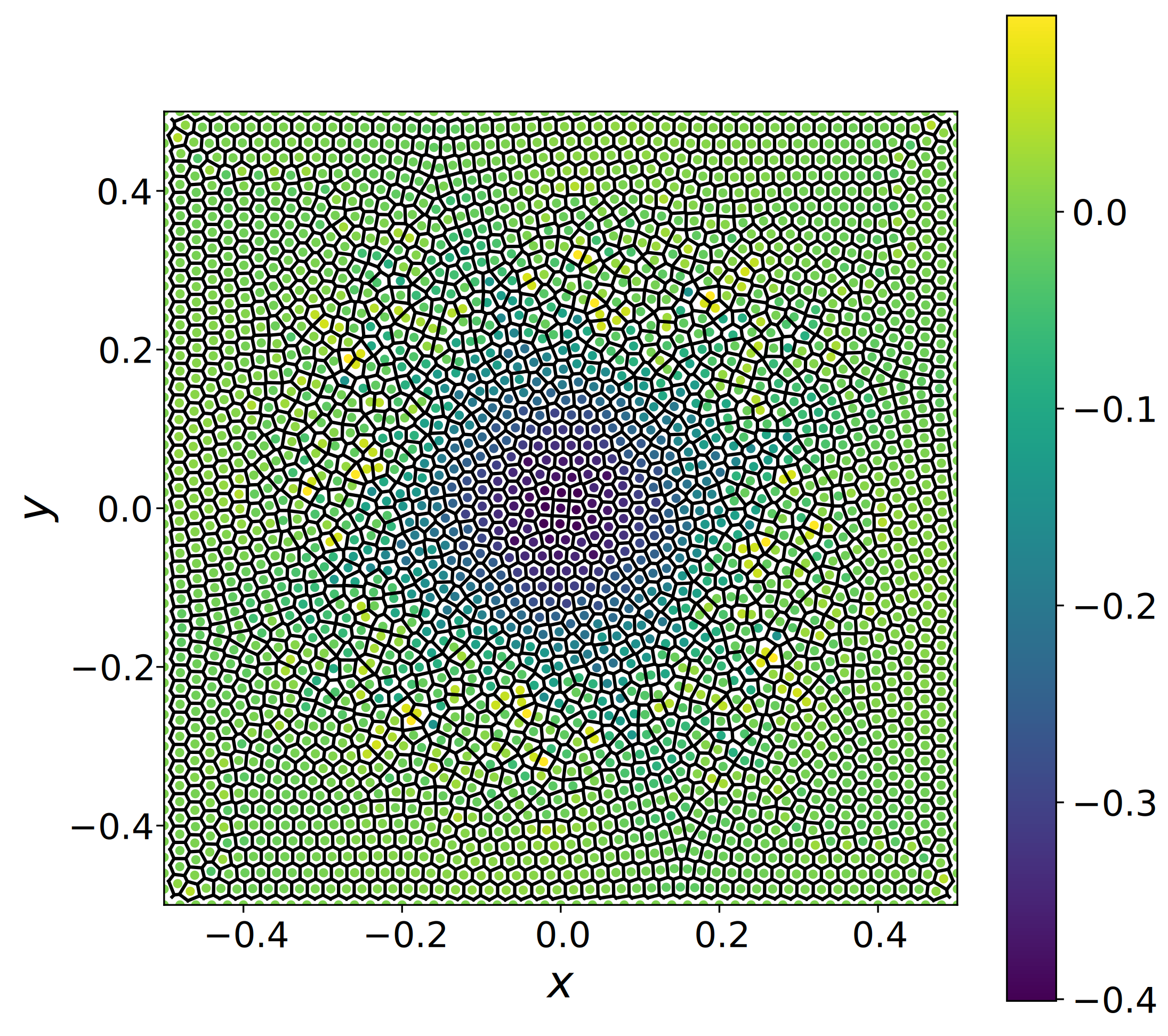} \hfil
	\includegraphics[
	width=0.45\linewidth]{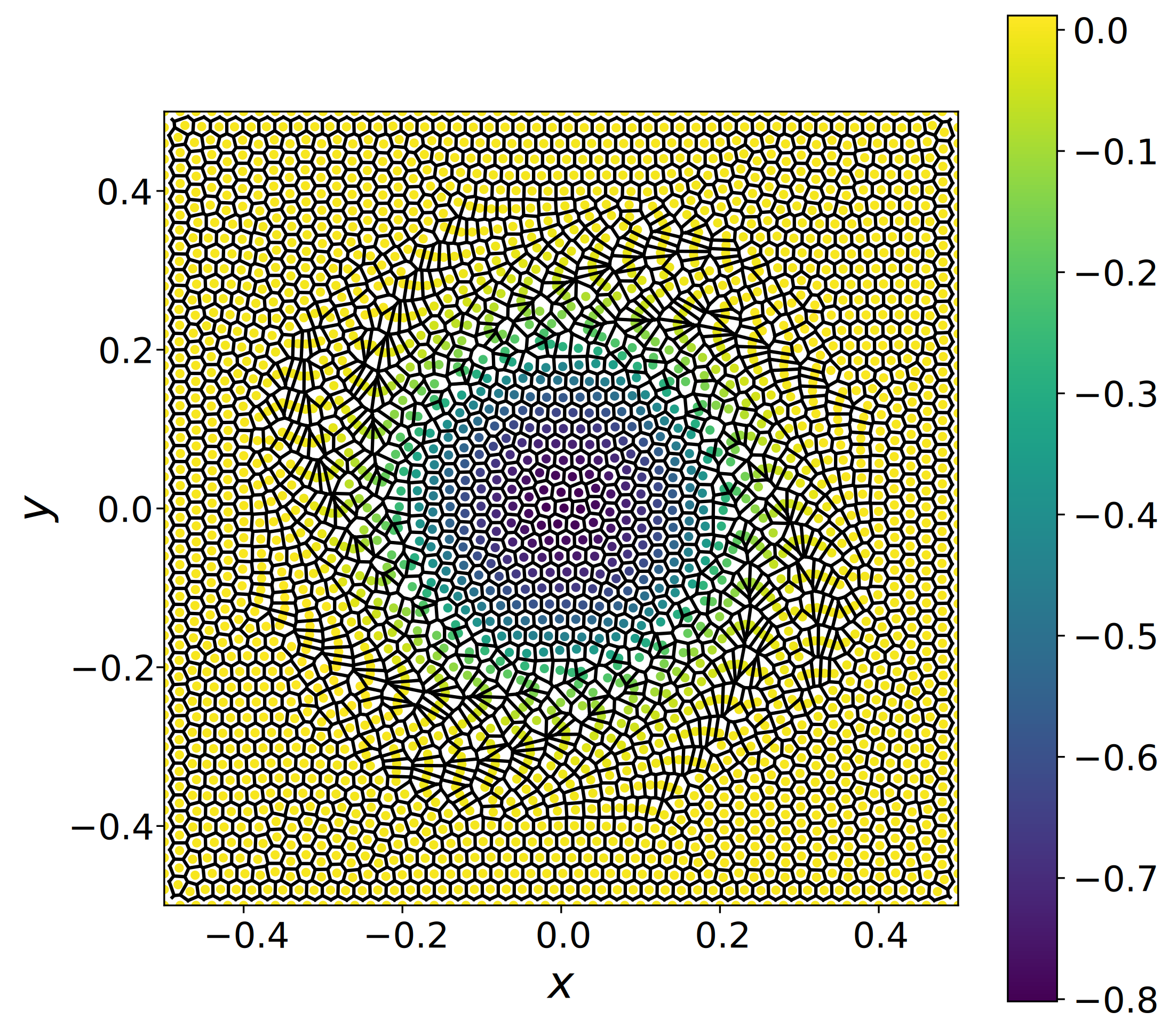} \\
	\includegraphics[
	width=0.32\linewidth]{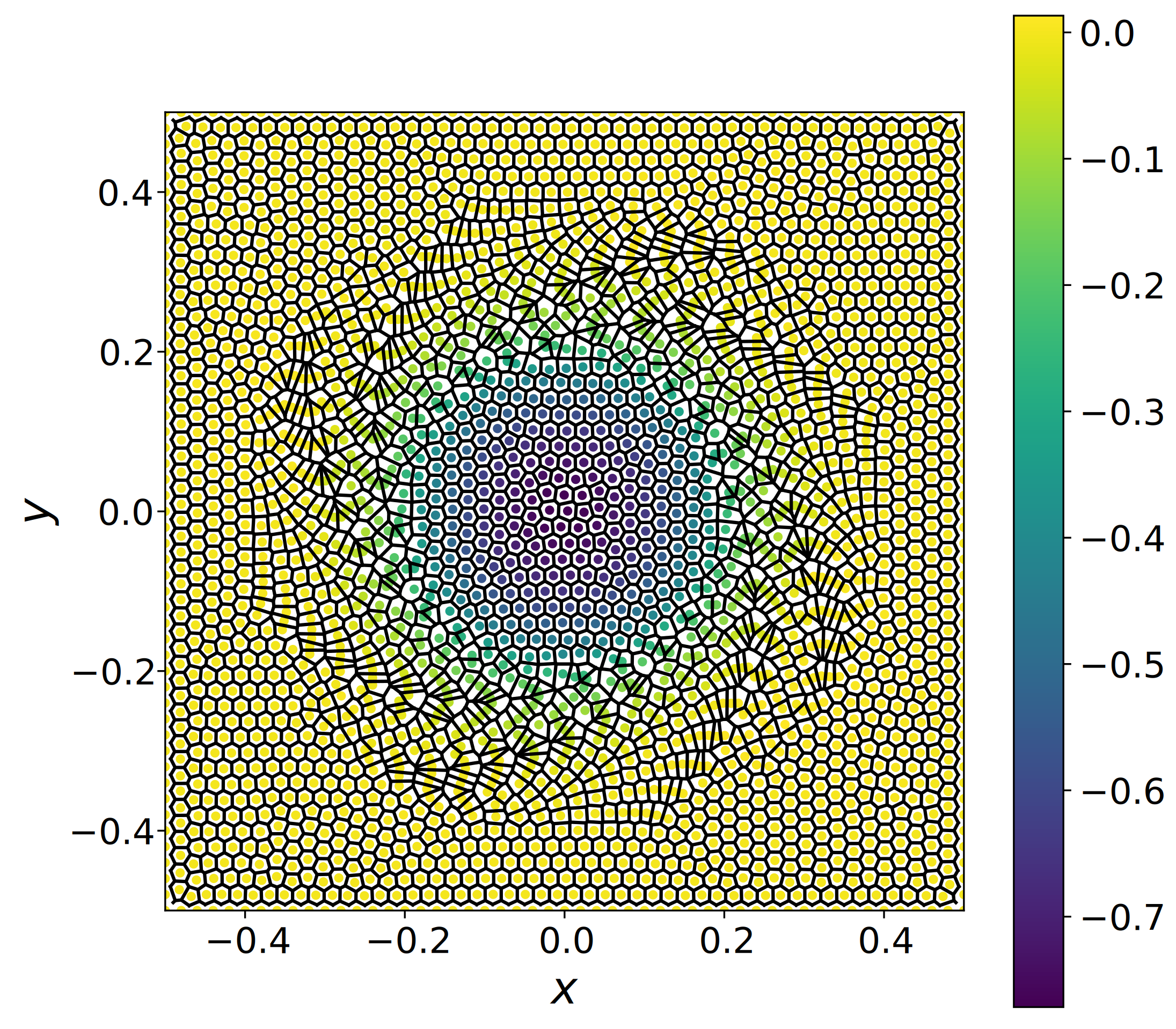}
	\caption{Snapshot of the system near $t=T/4$.
		Colors by pressure.
		Methods shown:
		Voronoi particles, de Goes' method, Gallouët and Mérigot, pFEM, and corrected pFEM.
	}
	\label{fig:snap_0.315}
\end{figure}

In Fig. \ref{fig:snap_0.315} snapshots of the system at time $t = 0.315$, very near one-quarter turn ($T/4 = 0.314$),  are shown. The simulation is acceptable in some ways, like a rather stable rotating core. However, even at this early time a number of undesirable features \deleted{can} have started to develop, like distorted shapes, with large aspect ratios and deviating a from centroidal diagram. Some particles are beginning to form pairs and chains. Also, as is easier to see in an animation, rapidly spinning cells appear. The pressure field is also seen to
\added{be inaccurate and to} reach values beyond its theoretical range.

\begin{figure}
	\centering
	\includegraphics[width=0.45\linewidth]{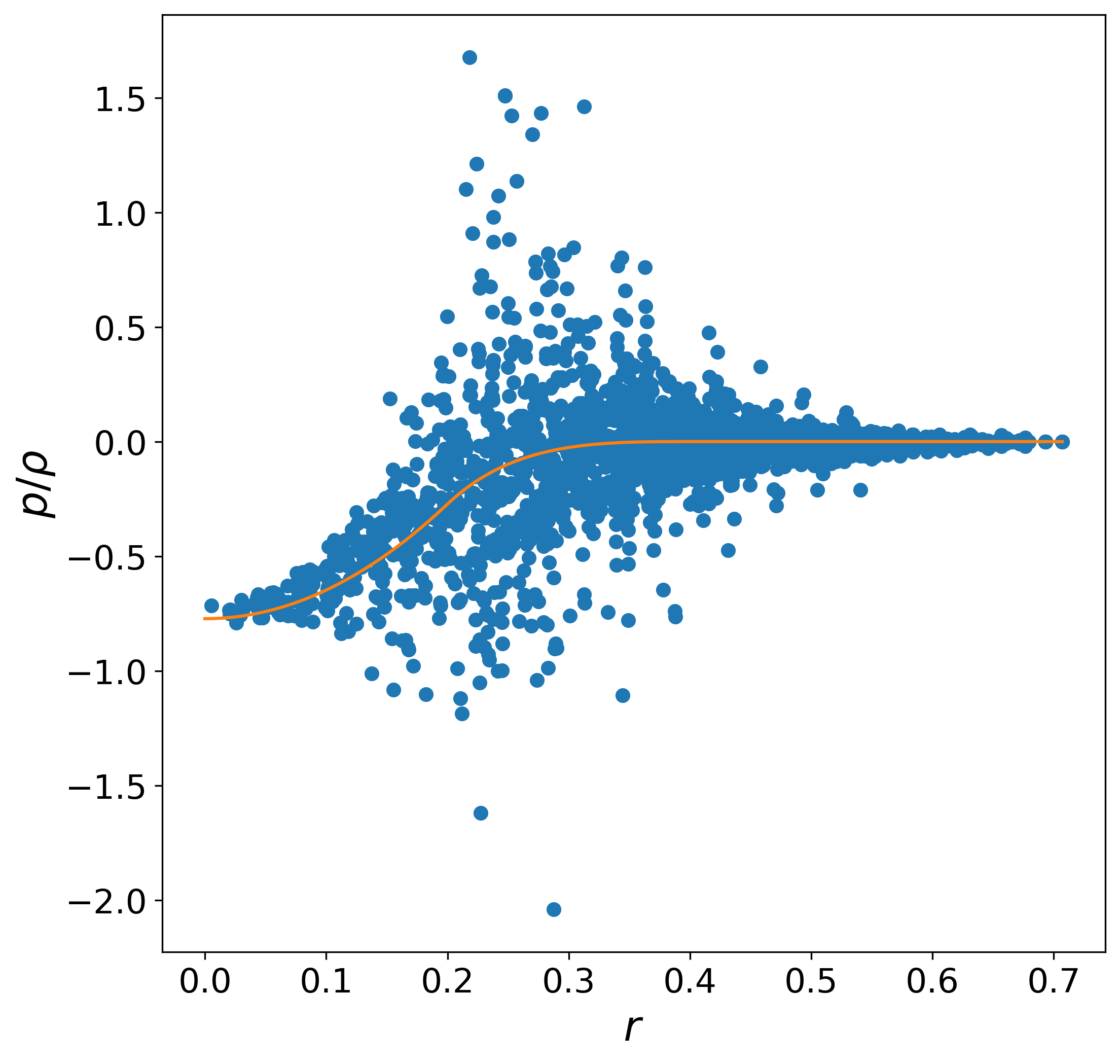} \hfil
	\includegraphics[width=0.45\linewidth]{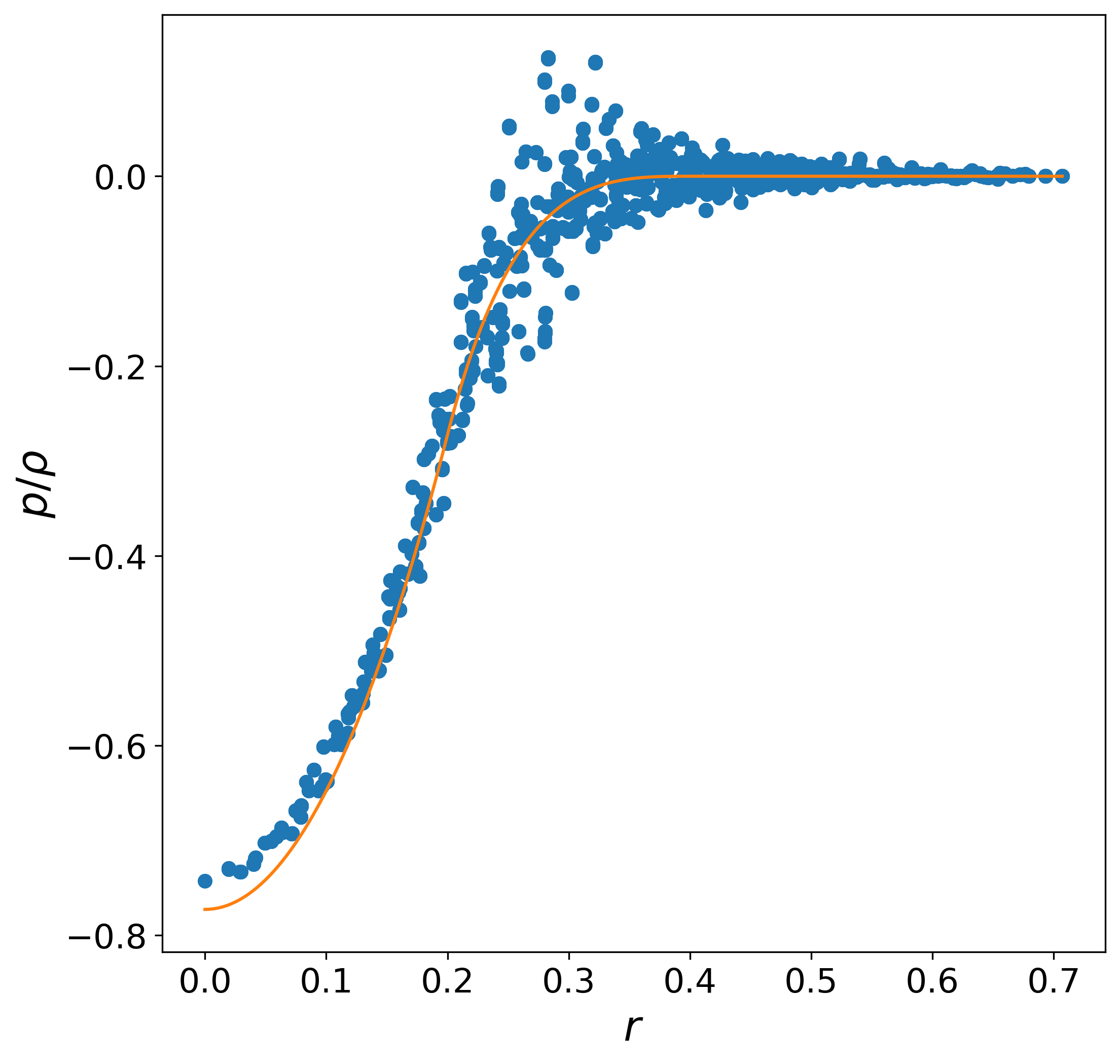} \\
	\includegraphics[width=0.45\linewidth]{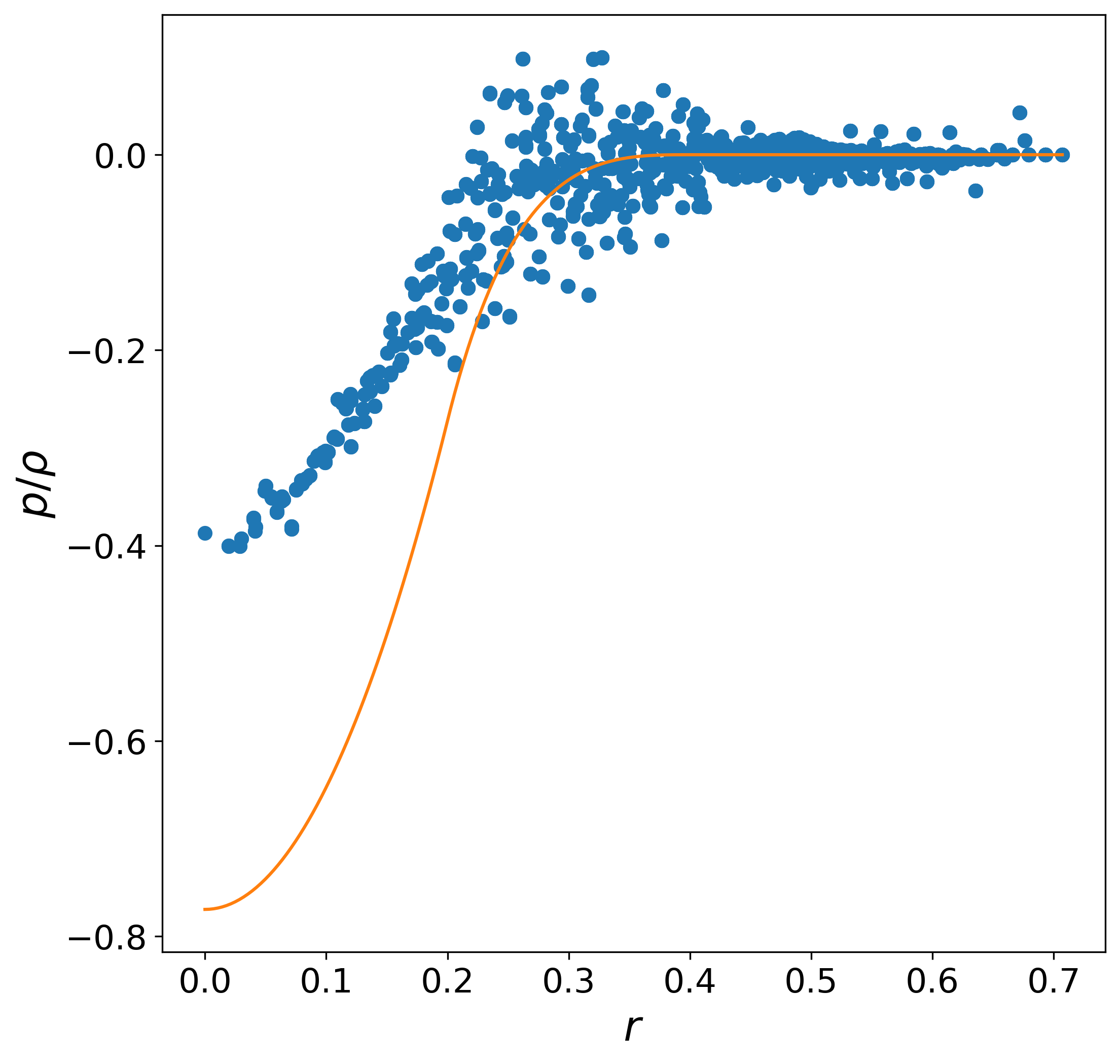} \hfil
	\includegraphics[width=0.45\linewidth]{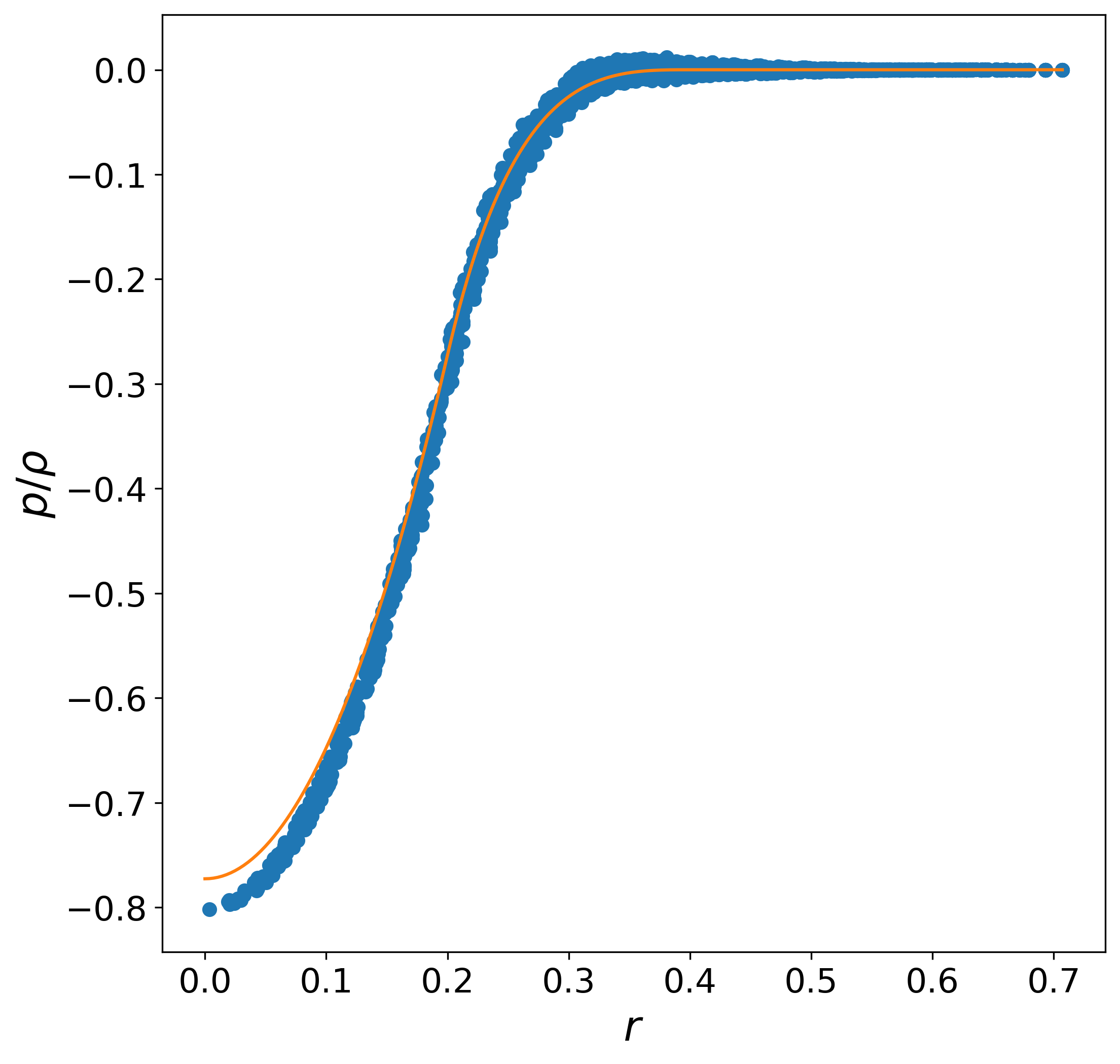} \\
	\includegraphics[width=0.45\linewidth]{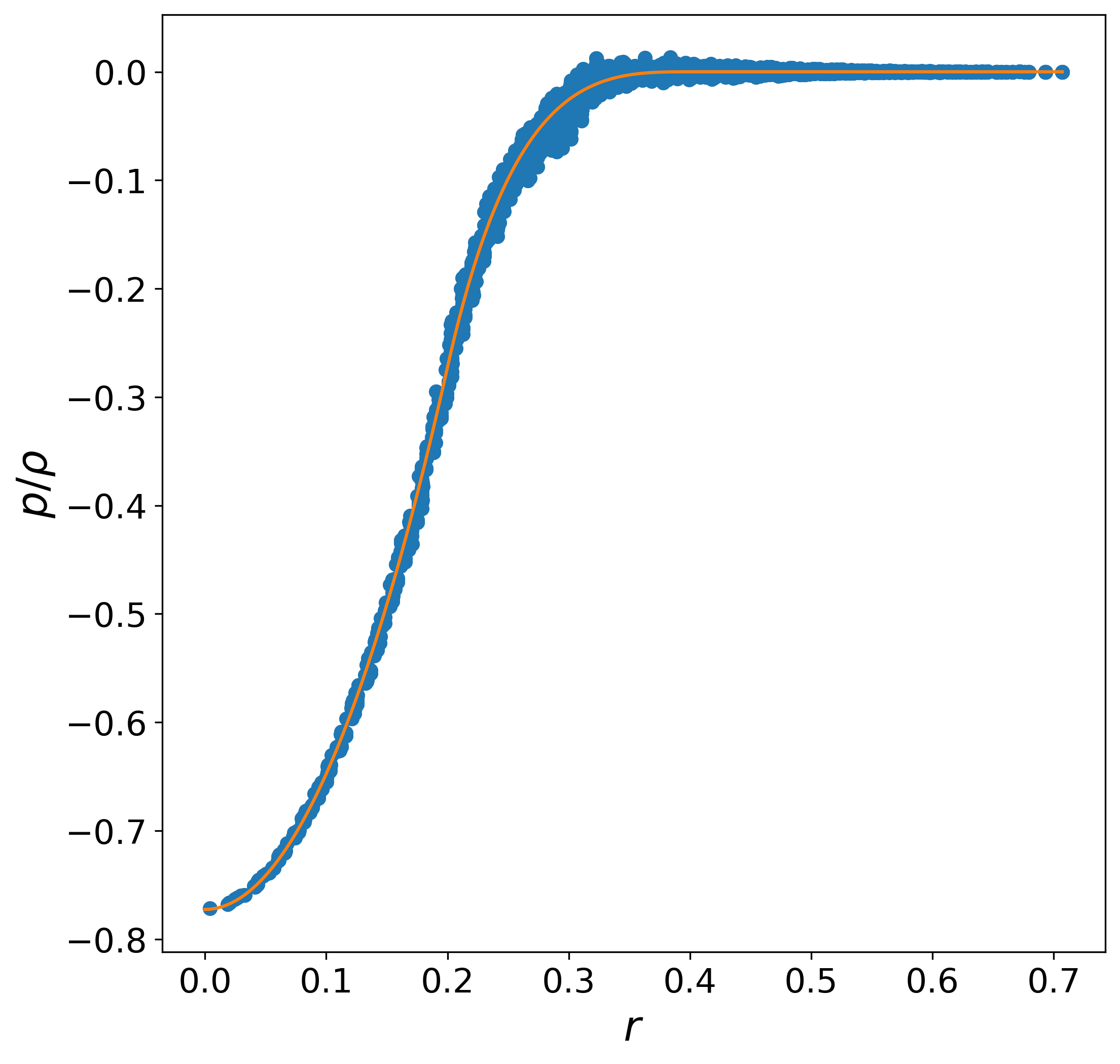}
	\caption{Pressure field. Orange lines: theoretical values.
		Shown:
		Voronoi particles, de Goes' method, Gallouët and Mérigot, pFEM, and corrected pFEM.
	}
	\label{fig:pressures}
\end{figure}

\begin{figure}
	\centering
	\includegraphics[width=0.45\linewidth]{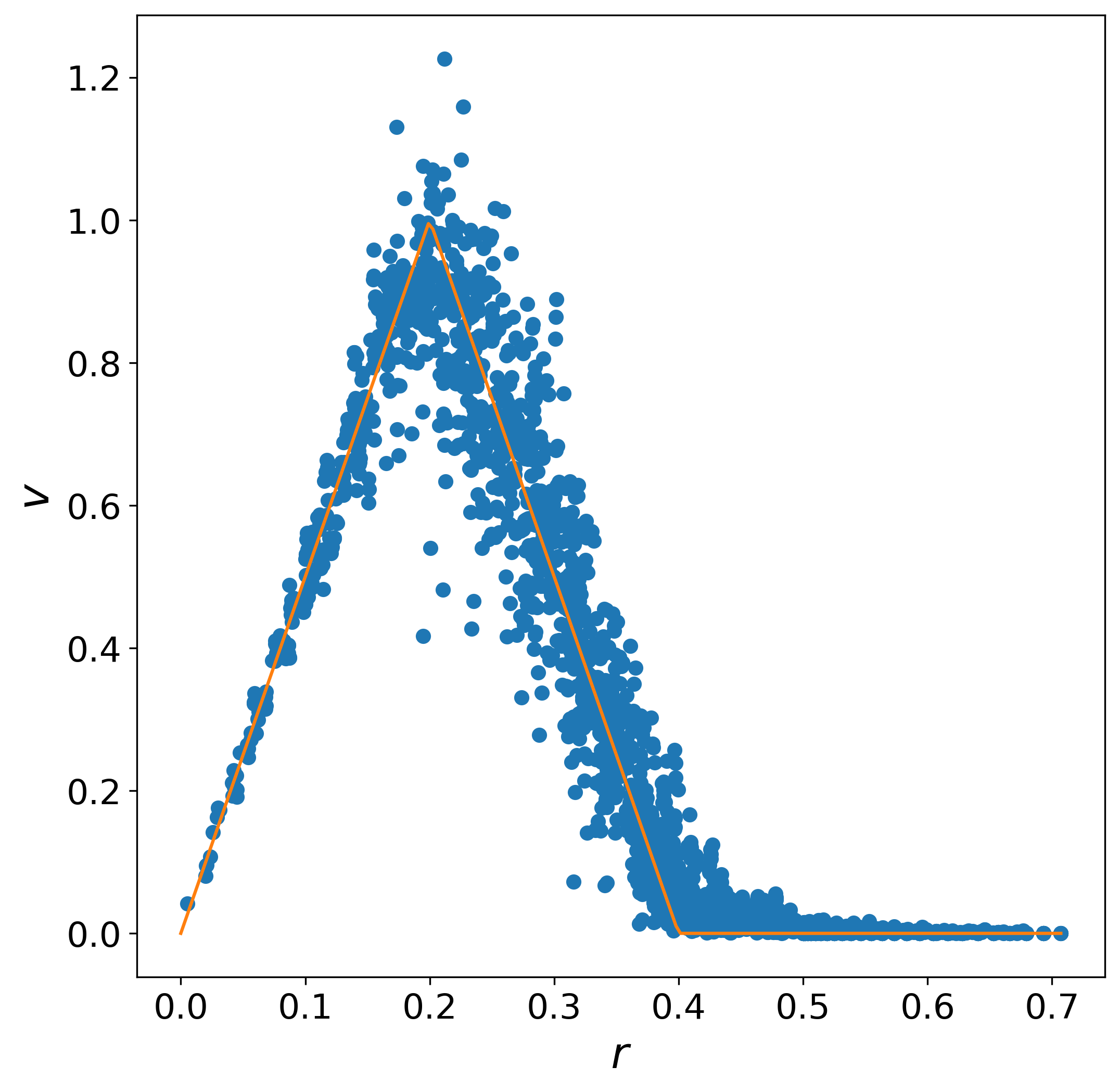} \hfil
	\includegraphics[width=0.45\linewidth]{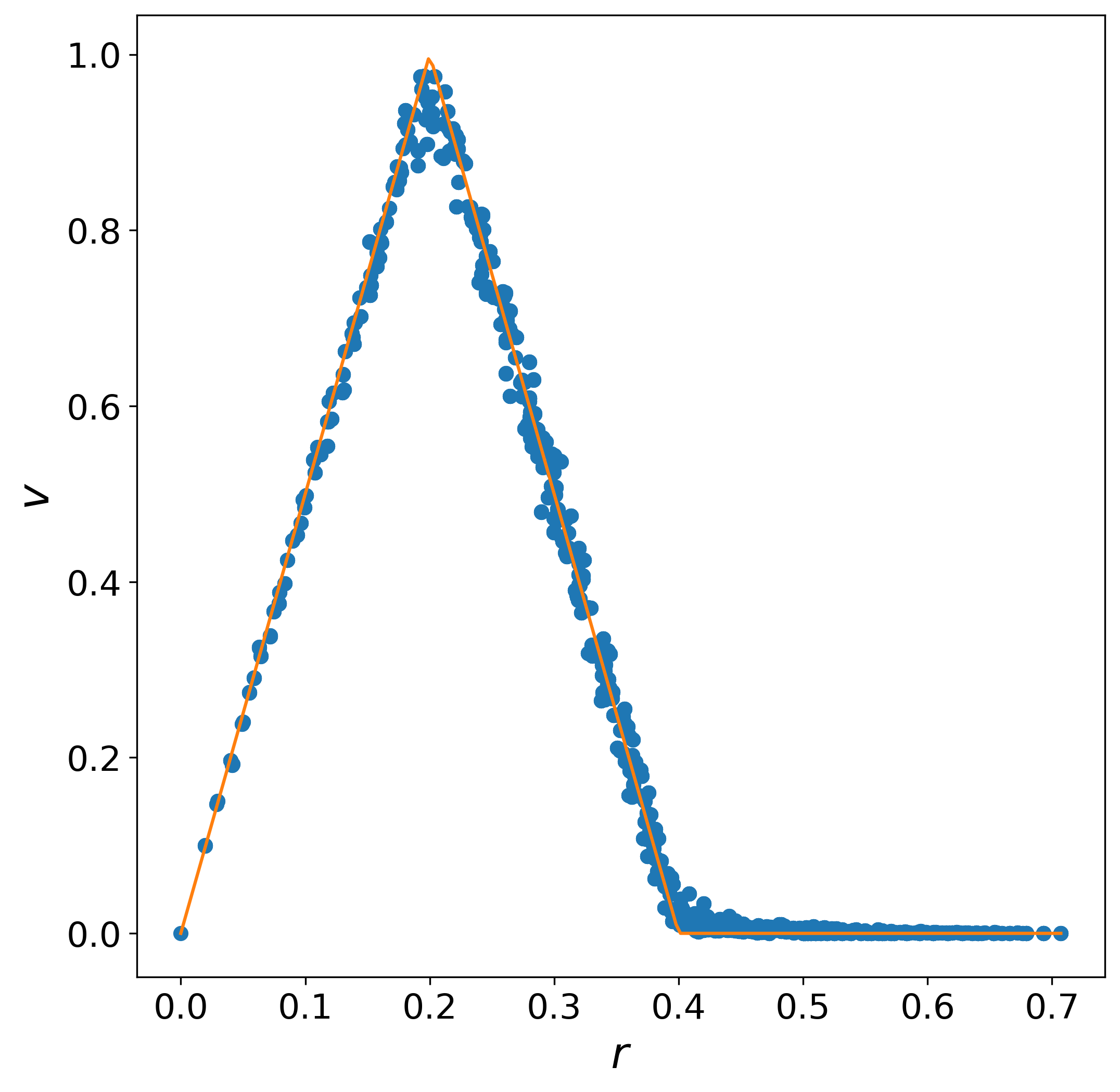} \\
	\includegraphics[width=0.45\linewidth]{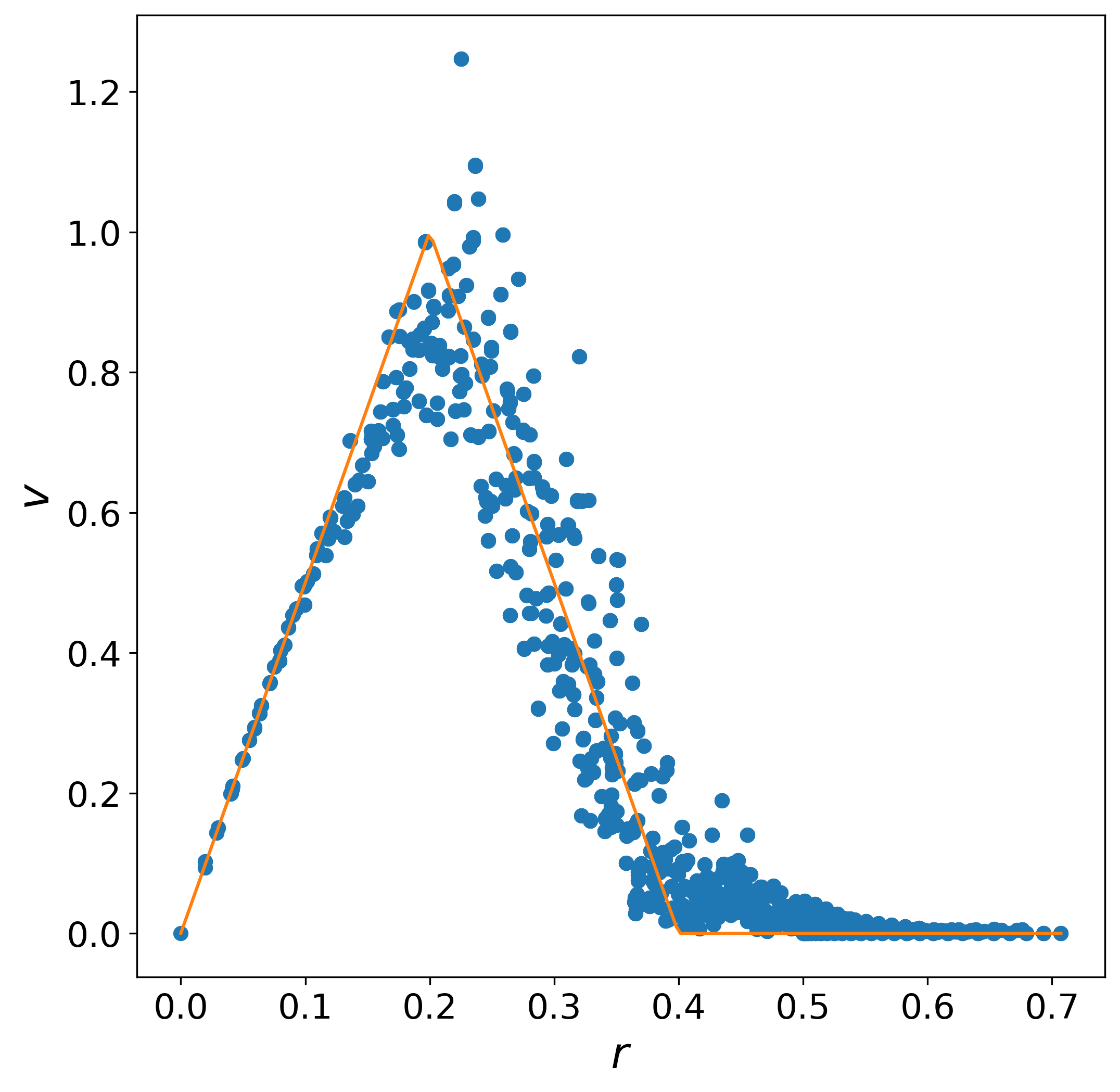} \hfil
	\includegraphics[width=0.45\linewidth]{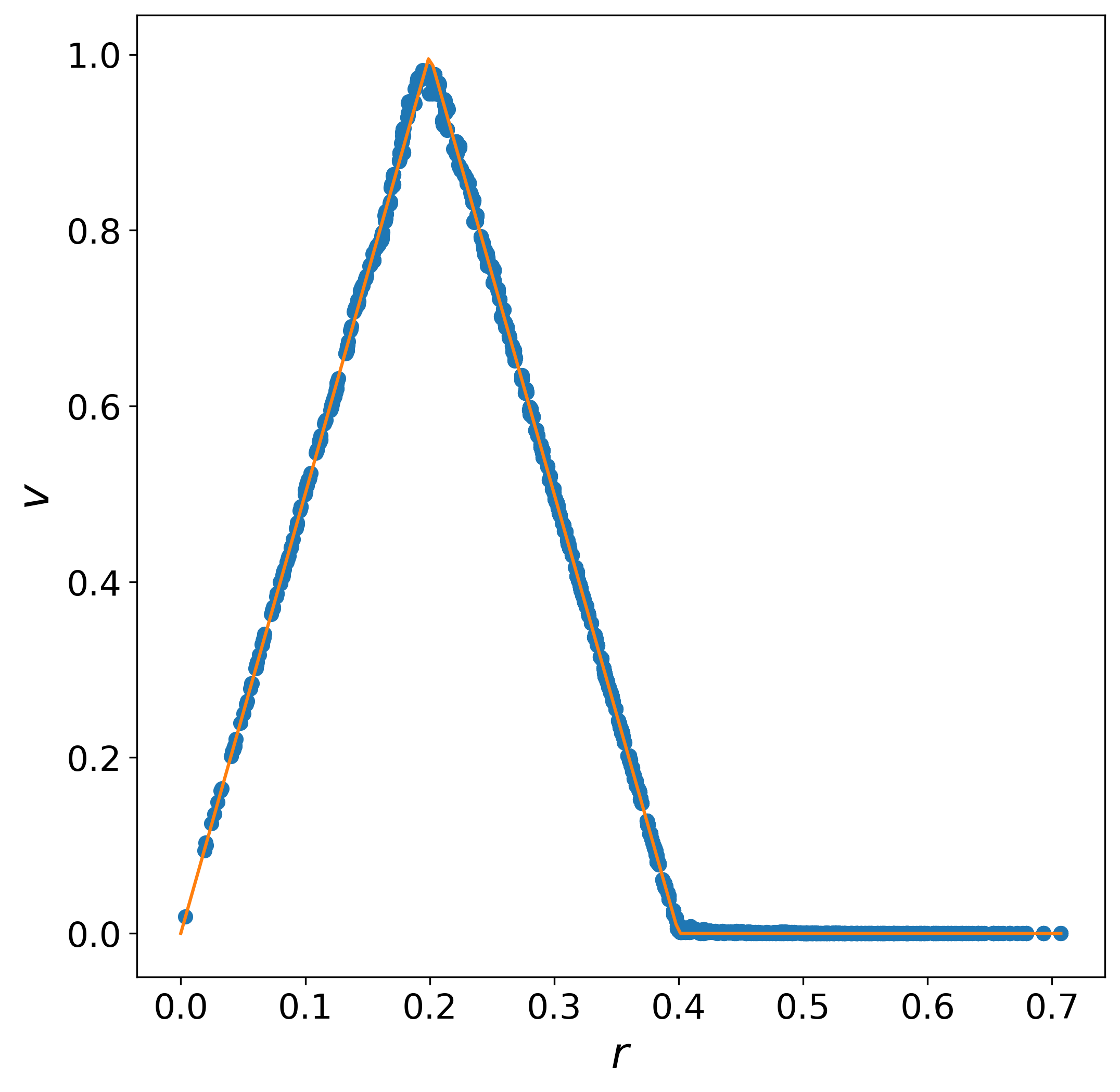} \\
	\includegraphics[width=0.45\linewidth]{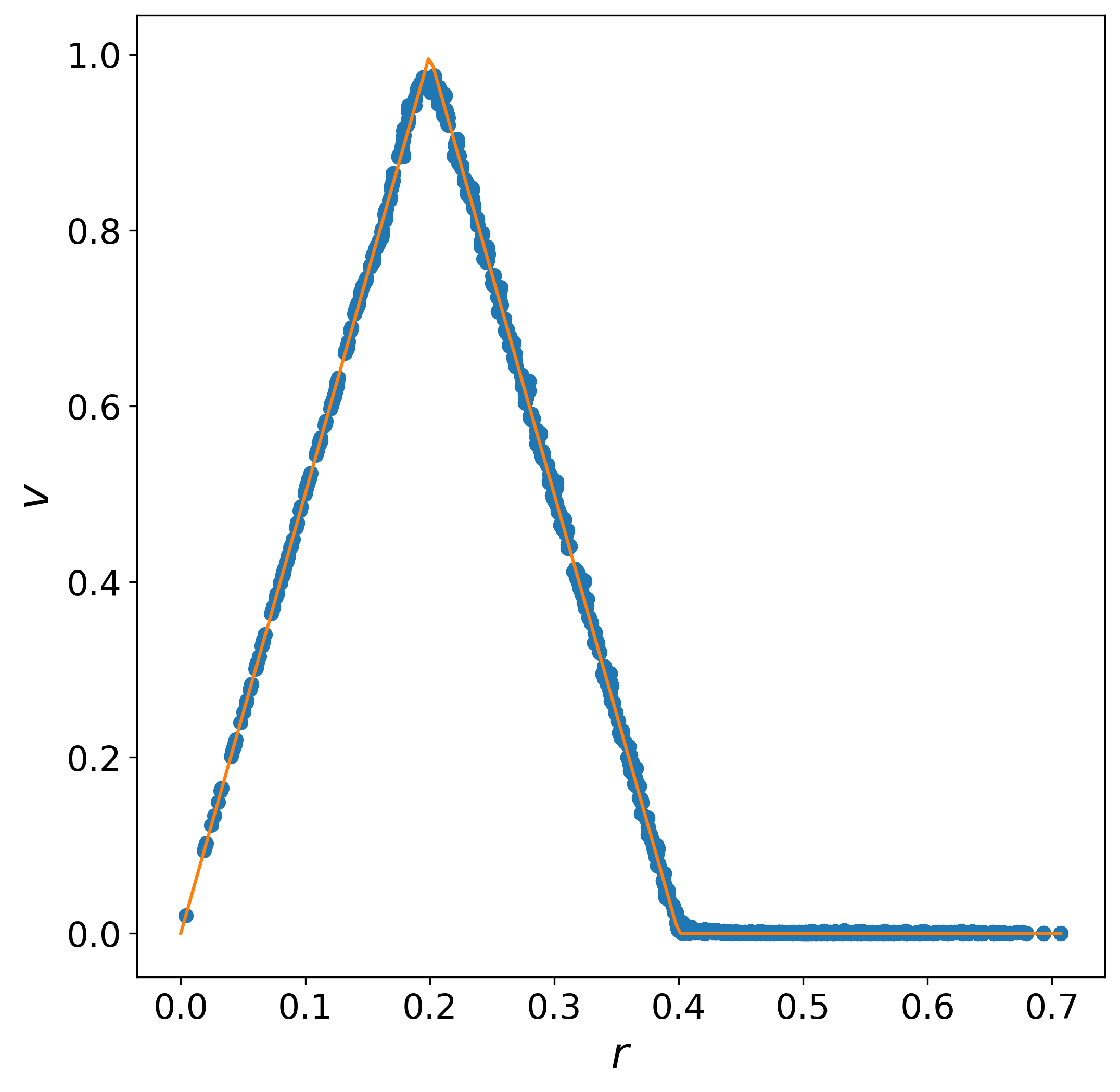} 
	\caption{Velocity modulus near $t=T/4$. Orange lines: theoretical values.
		Shown:
Voronoi particles, de Goes' method, Gallouët and Mérigot, pFEM, and corrected pFEM.
	}
	\label{fig:velocities}
\end{figure}

This is clearer in the pressure field is shown in Fig. \ref{fig:pressures}, with somewhat reasonable results, especially in the core, but very large dispersion in the corona region. The velocity field, Fig. \ref{fig:velocities} is smoothed out from its initial piece-wise linear shape, but no large anomalies are found at this stage.

\begin{figure}
	\centering
	\includegraphics[width=0.45\linewidth]{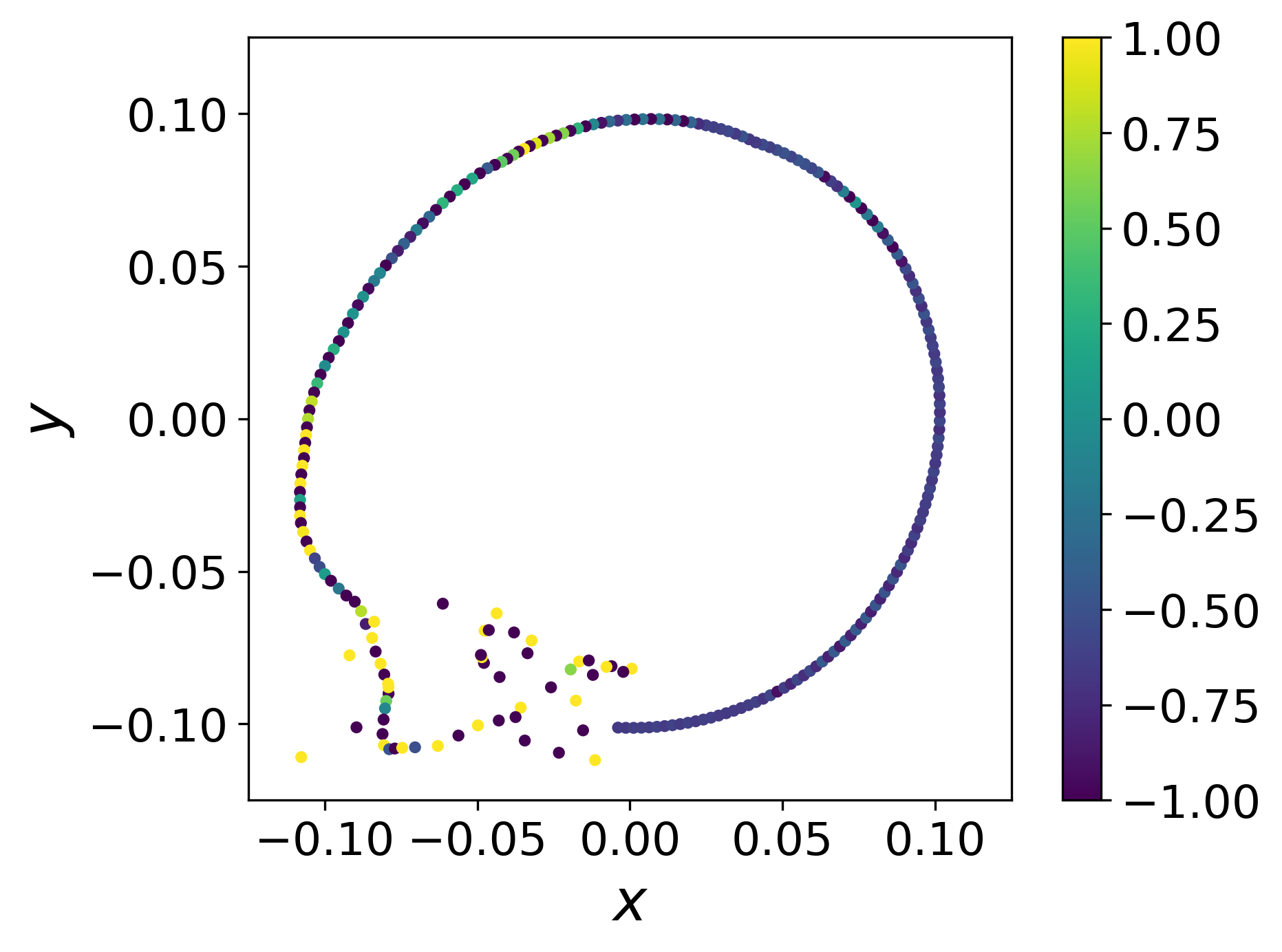} \hfil
	\includegraphics[width=0.45\linewidth]{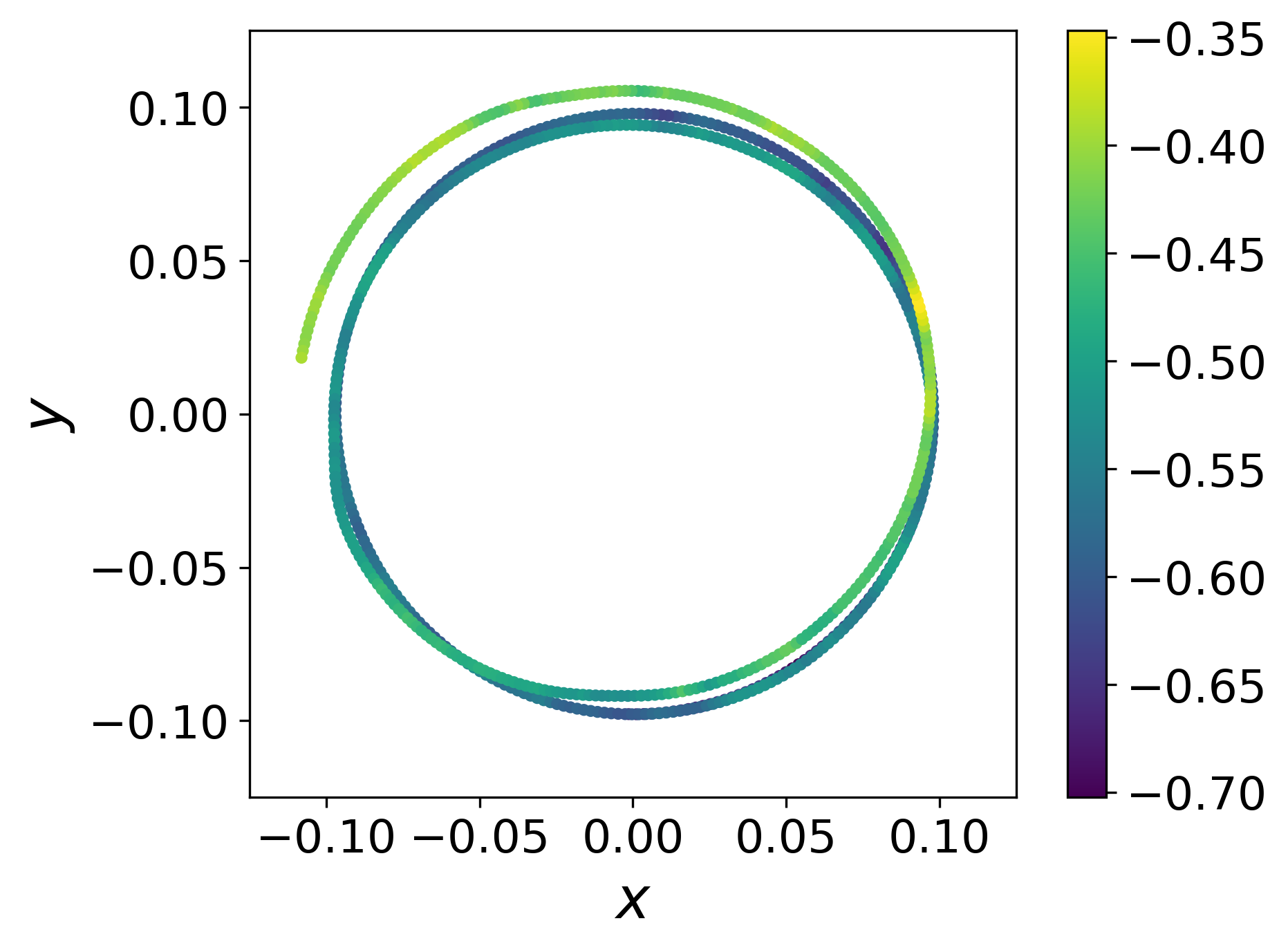} \\
	\includegraphics[width=0.45\linewidth]{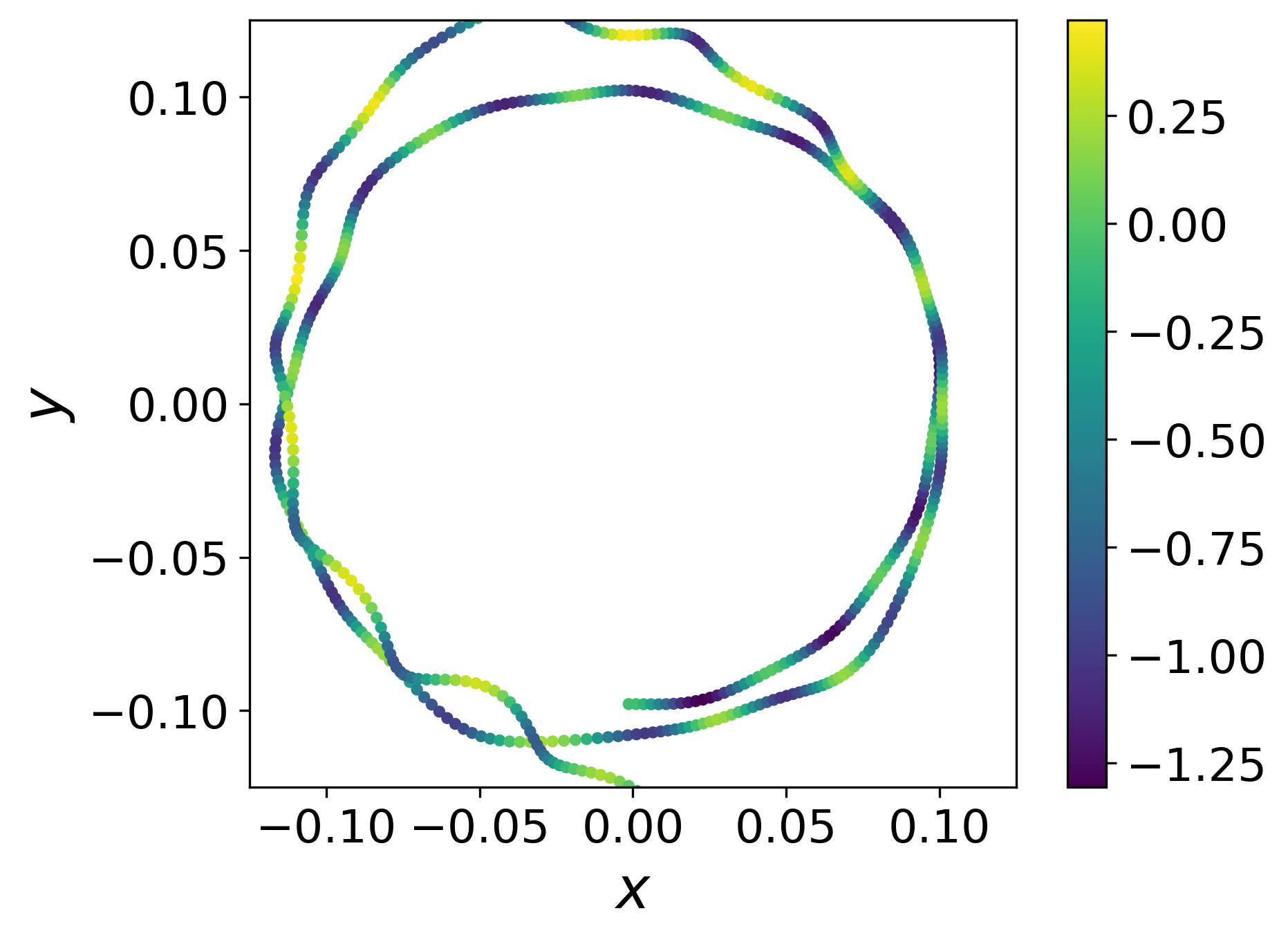} \hfil
	\includegraphics[width=0.45\linewidth]{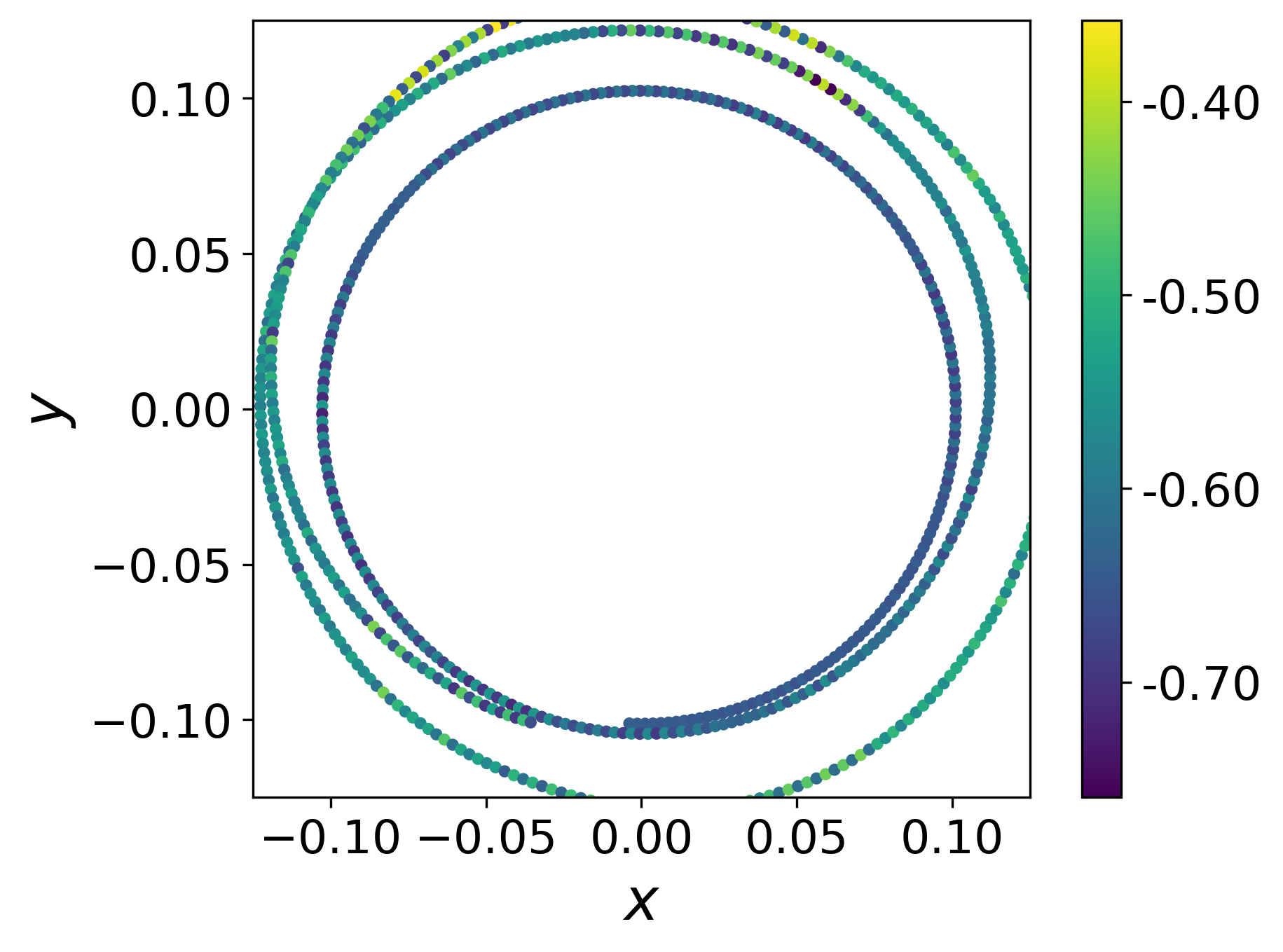} \\
	\includegraphics[width=0.45\linewidth]{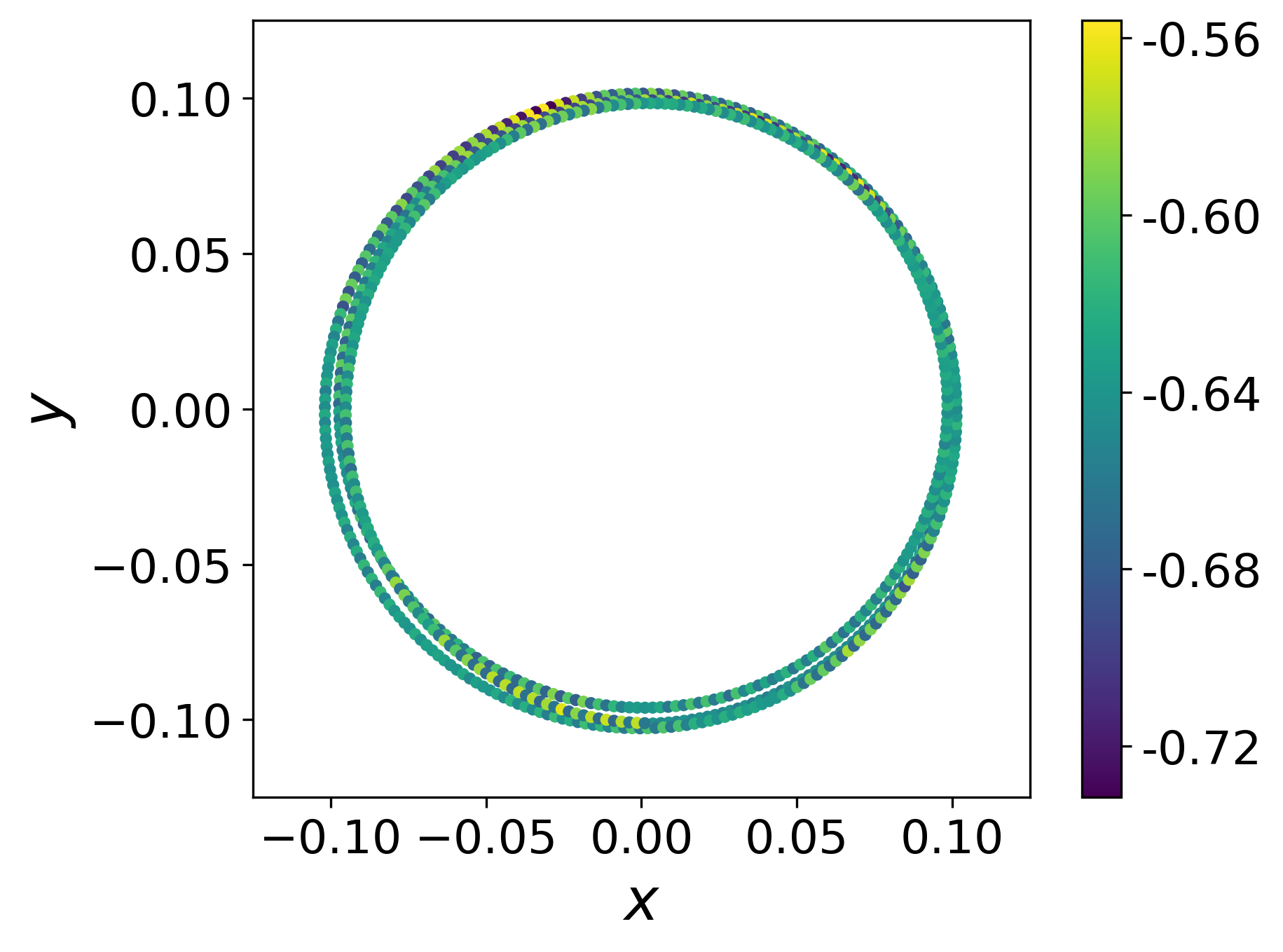}

	\caption{Trajectory of a particle starting in the core region.
		Colors by pressure.
		Shown:
Voronoi particles, de Goes' method, Gallouët and Mérigot, pFEM, and corrected pFEM.
	}
	\label{fig:traj}
\end{figure}

In Fig. \ref{fig:traj}, the path of a particle belonging to the core region is shown. It should ideally perform three whole circular turns, but its trajectory deviates from a circle after about a half turn. More seriously, the simulation fails around $T$. Its pressure should also be constant, as it nearly is in the beginning. Then, oscillations appear, which become more severe before the failure.

\begin{figure}
	\centering
	\includegraphics[width=0.75\linewidth]{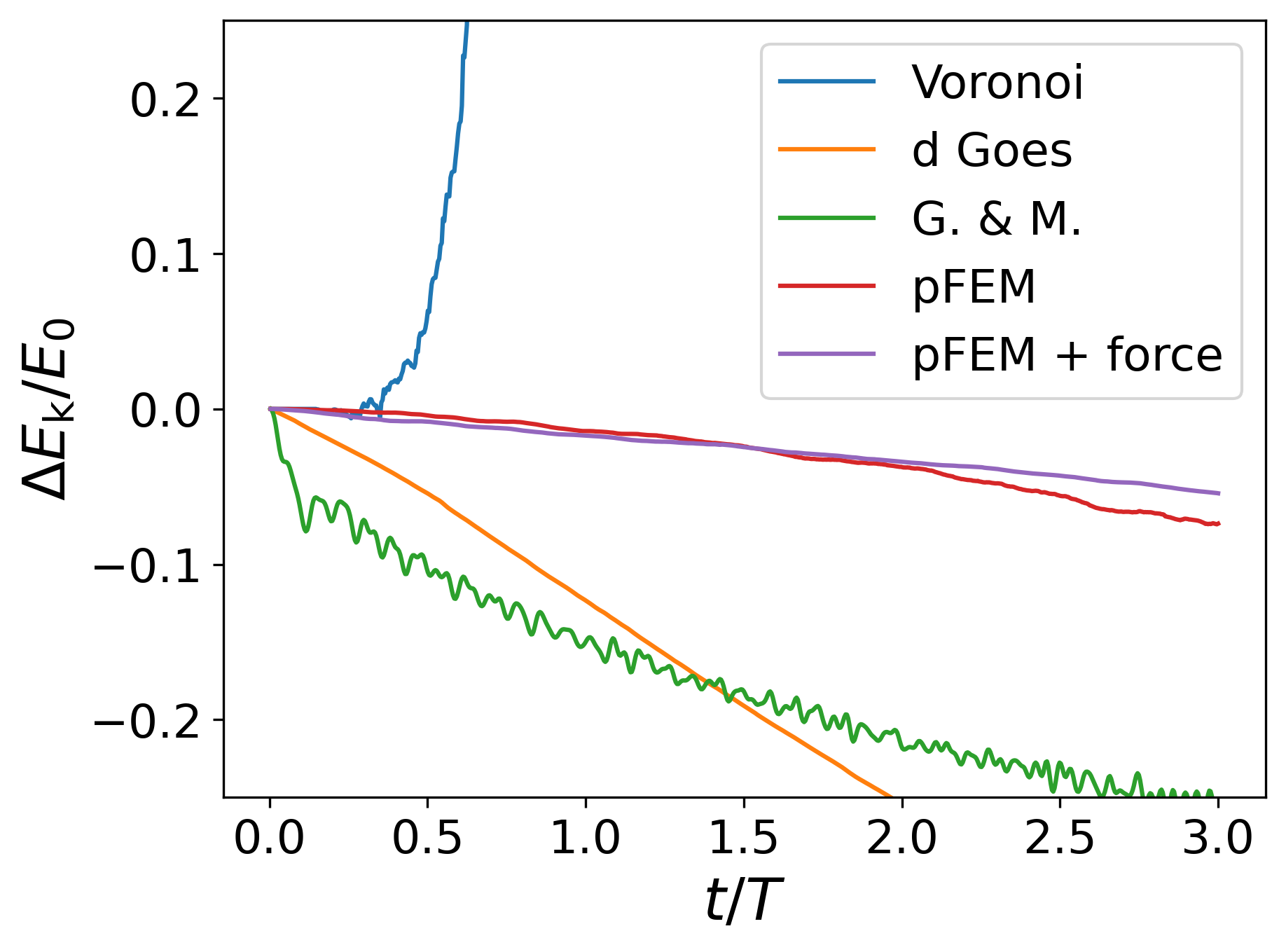} \\
	\includegraphics[width=0.75\linewidth]{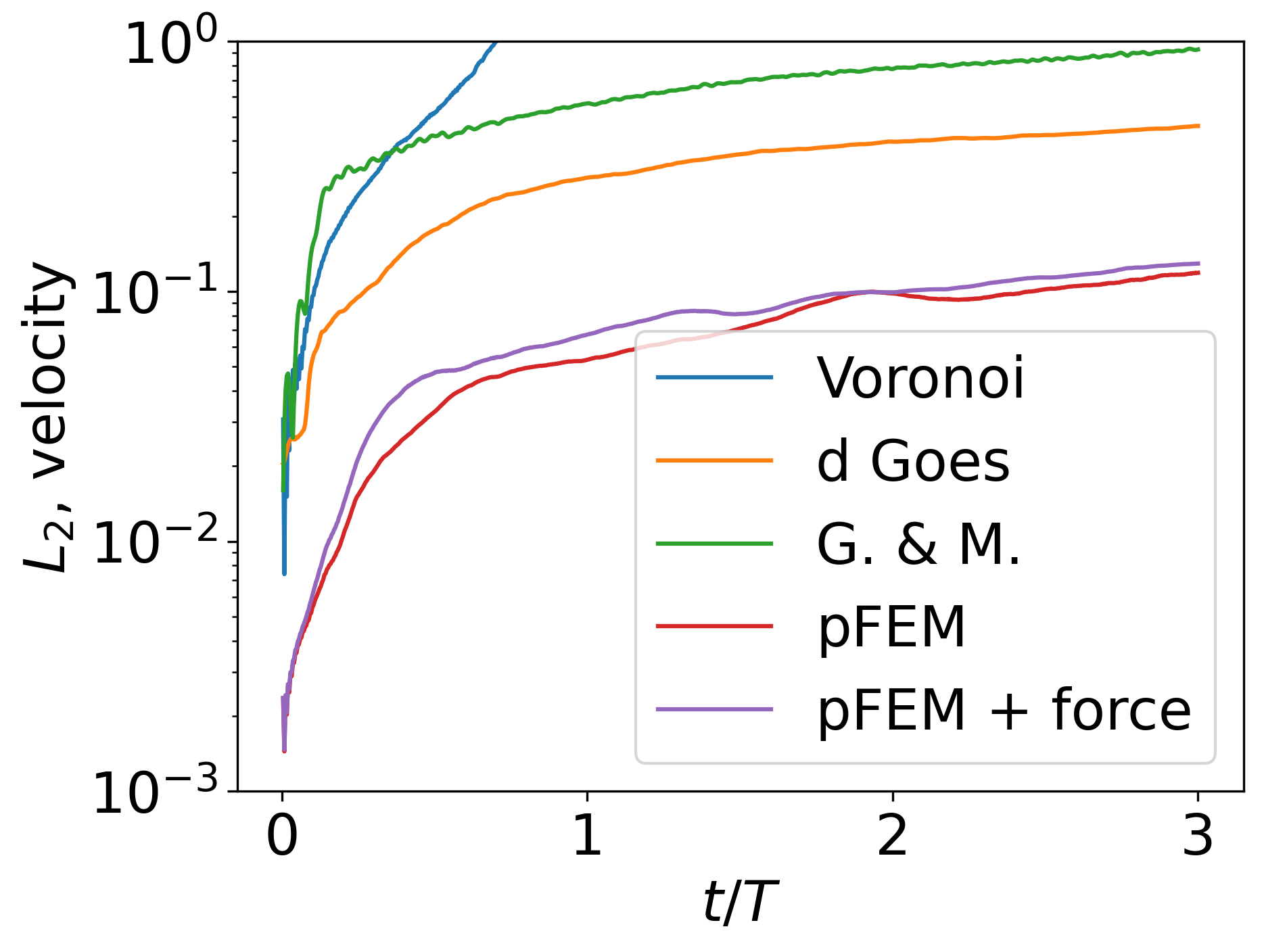} \\
	\includegraphics[width=0.75\linewidth]{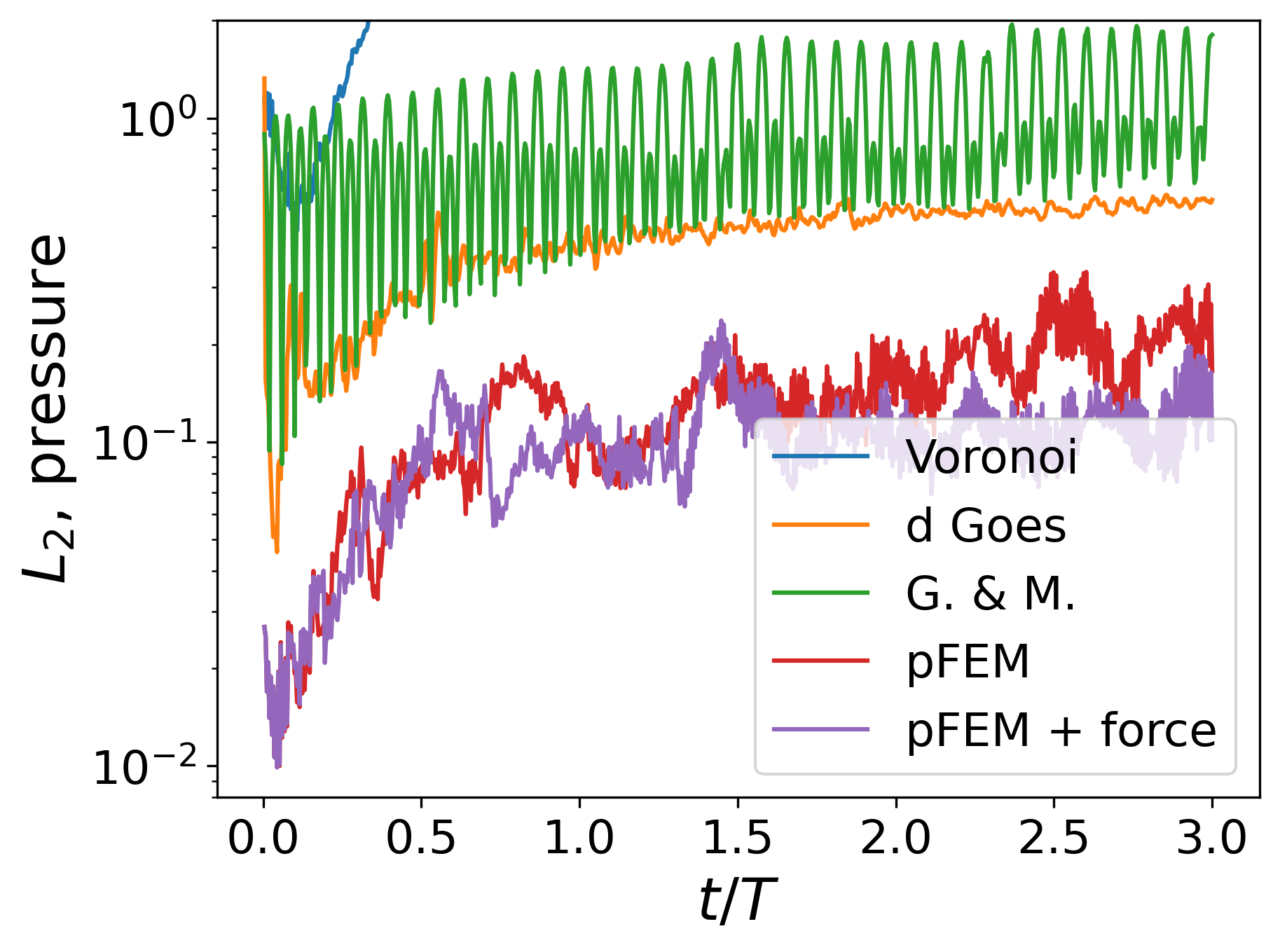}

	\caption{
		Gresho vortex simulations.
		Left: change in the relative value of the kinetic energy.
		Middle: relative $L_2$ distance between the velocity field and its theoretical value.
		Right: relative $L_2$ distance between the pressure field and its theoretical value.}
	\label{fig:ek_L2}
\end{figure}

As a way to monitor the quality of the simulation, we may track the time evolution of
the kinetic energy, $E_\text{k} := (\rho/2) \sum_i  V_i u_i^2$. It is more convenient to plot its relative change from its initial value:
$( E_\text{k}(t) - E_\text{k}(0) )/ E_\text{k}(0) $. Fig. \ref{fig:ek_L2} (left) shows an increase in this quantity, and hence spurious energy creation, which blows up before one turn.

We may also plot the $L_2$ distance between our field and the theoretical one:
\[
L_2:= \frac{\sum_i |\bfu_i - \bfu(\bfr_i)|^2 - \sum_i u(\bfr_i)^2}{\sum_i u(\bfr_i)^2} ,
\]
where $\bfu(\bfr_i)$ means the theoretical value of the Gresho velocity field at the position of particle $i$.
This difference is depicted in Fig. \ref{fig:ek_L2} (middle), showing a rapid degradation at the early stages of the simulation.

The accuracy of the pressure field may be tested by the same $L_2$ measure (adapted for scalars), and shown in Fig. \ref{fig:ek_L2}. It starts quite high and rather constant, and then it diverges.

\subsubsection{Power particles}
\label{ss:simulations:DG}

Given its importance in the recent development of methods based on power diagrams, we include a dG simulation.
It starts from a centroidal equal-volume power
diagram, which is obtained at the beginning of the simulation by a Lloyds-like procedure.
The time integration is explicit, and consists of volume equalization as in Eq.~\eqref{eq:NR}, followed by an evaluation of the pressure from the PPE as in Eq. \eqref{eq:PPE} to get the pressure field.
Particles are moved, including the gradient of this pressure field, from the centroid of their cells. 

Results \added{in} Fig. \ref{fig:snap_0.315} show a well ordered diagram (to be expected, since particles are not allowed to go far away from the centroids). The trajectory in Fig. \ref{fig:traj} shows a rather circular trajectory. The particle, however, fails to return to its initial location. This can be attributed to dissipation of velocity. Indeed, Fig. \ref{fig:ek_L2} (left) shows that kinetic energy is lost, up to about $25\%$ at the end of a time corresponding to three rotations. An overall decrease in the velocity field can also be observed in Fig. \ref{fig:velocities}. The energy dissipation may be due to the fact that the dynamics are not entirely consistent: each particle is moved from its corresponding centroid, but keeping its velocity value intact. We suspect (but have not tested) that an interpolation procedure should be applied, in order to better guess the velocity field at the centroid (see Ref.~\cite{duque_assignment_2020} for a discussion of interpolating procedures.) In Fig. \ref{fig:ek_L2} (middle), this velocity field is also seen to deteriorate at the early stages of the simulation, even though the rate is reduced afterwards.
Meanwhile, the pressure field, Fig. \ref{fig:ek_L2}, clearly deviates from its theoretical value, even if it does not grow much.

. 

\subsubsection{Gallouët and Mérigot}
\label{ss:simulations:GM}

In the GM method, the setup is similar to the de Goes simulation just described.
The main difference is the appearance of a spring-like force between each particle and the centroid of its cell, and the vanishing of any other force (no pressure gradient is used.)
Results are shown for a spring period $\tau :=2\pi/\omega = 0.1 $.
In their original work, these authors quote this parameter as a multiple of $\Delta t$ (in this case, $ \tau = 20\Delta t $ ). However, in our opinion, this parameter should be considered problem-dependent. In this case, each spring would oscillate about 10 times per turn. Of course, $\Delta t$ should still be small enough so that each spring oscillation is correctly resolved, with a minimum value of about $\tau / (\Delta t) > 10$.


In Fig. \ref{fig:snap_0.315}, the snapshot shows nodes that are close to centroidal, as enforced by the spring force. The trajectory of Fig. \ref{fig:traj} clearly shows oscillations due to the springs, both in the trajectory itself as in the pressure value. The particle comes back close to its initial position after one turn, meanders on its second turn, and deviates after that. The pressure field of Fig. \ref{fig:pressures} is removed from its theoretical value. However, at variance with the power particle method, this field is seen to oscillate up and down about the theoretical value. Notice this pressure is entirely given by the weights, as in Eq. \eqref{eq:GM_pressure_weight}. In this sense, the pressure profile in Fig.~\ref{fig:pressures} is caught in a moment when it is overall too high, but it actually fluctuates above and below the theoretical function. The velocity field Fig. \ref{fig:velocities} is very noisy, and worse than the Voronoi particles result. Nevertheless, at variance with the latter, the simulation does not fail.

The kinetic energy, Fig. \ref{fig:ek_L2} (left), presents oscillations due to the springs, and a high dissipation rate, comparable to the dG method. Meanwhile, Fig. \ref{fig:ek_L2} (middle) a similar deterioration of the velocity field, with an enhanced deterioration at the early stages, and a similar rate afterwards. The pressure $L_2$ deviation, in Fig. \ref{fig:ek_L2}, shows very clear oscillations, and is quite inaccurate.

\subsubsection{Standard pFEM}
\label{ss:simulations:pFEM}

Results from this method are much better, e.g. Fig. \ref{fig:snap_0.315} shows a snapshot that is much more orderly, specially in the
core region --- however, elongated cells are still present. The trajectory in Fig. \ref{fig:traj} is close to circular, and ends up near the original location after one turn, deviates on the second, then comes back on the third. The pressure is rather stable, even though oscillations begin to occur after approximately one turn.

The pressure field of Fig. \ref{fig:pressures} is very satisfactory (except at the inner core, where it is slightly high), and so is the velocity field in Fig. \ref{fig:velocities}.

Finally, Fig. \ref{fig:ek_L2} (left) shows very little energy dissipation, while Fig. \ref{fig:ek_L2}  (middle) confirms the velocity profile degrades at a slow rate, after an initial regime of rapid deterioration.
Fig. \ref{fig:ek_L2} shows an accurate pressure field, with limited growth.













\subsubsection{Corrected  pFEM}
\label{ss:simulations:cpFEM}

Finally, we show simulations including the newly derived force. The adjustable parameter $\Pi \beta$ can be set to any value. One too low will of course produce the same result as standard pFEM. One too high results in nonphysical results, much like what happens with the spring parameter in the GM case. We have chosen a value of $\tau = 2\pi /\omega = 2.5 T$, so that the influence of the new term should only make a clear difference after one turn or so. This is a remarkable deviation from the value used in the GM simulation, but here the force is to be seen as a small correction to quite accurate pressure forces, whereas in GM the spring force was the only one present.

Fig. \ref{fig:snap_0.315} shows a snapshot that is almost the same as in the pFEM simulation, if slightly more orderly. The velocity profile of Fig.~\ref{fig:velocities} looks also almost identical. At this moment, the additional forces have not had the time to alter the simulation much. The pressure field of Fig. \ref{fig:pressures} is more accurate in the inner core, but shows
\replaced{greater}{larger} dispersion in the corona.

It seems this procedure is somewhat superior, as shown in the trajectory in Fig. \ref{fig:traj}, that stays close to circular for the three turns. Pressure oscillations do occur, but are limited in range.

Finally, Fig. \ref{fig:ek_L2} confirms that both pFEM simulations are rather similar, but the plain one performs better at earlier times, with a crossover from about one period and a half, where the corrected pFEM slightly outperforms the other method.
%

\begin{figure}
	\centering
	\includegraphics[width=0.75\linewidth]{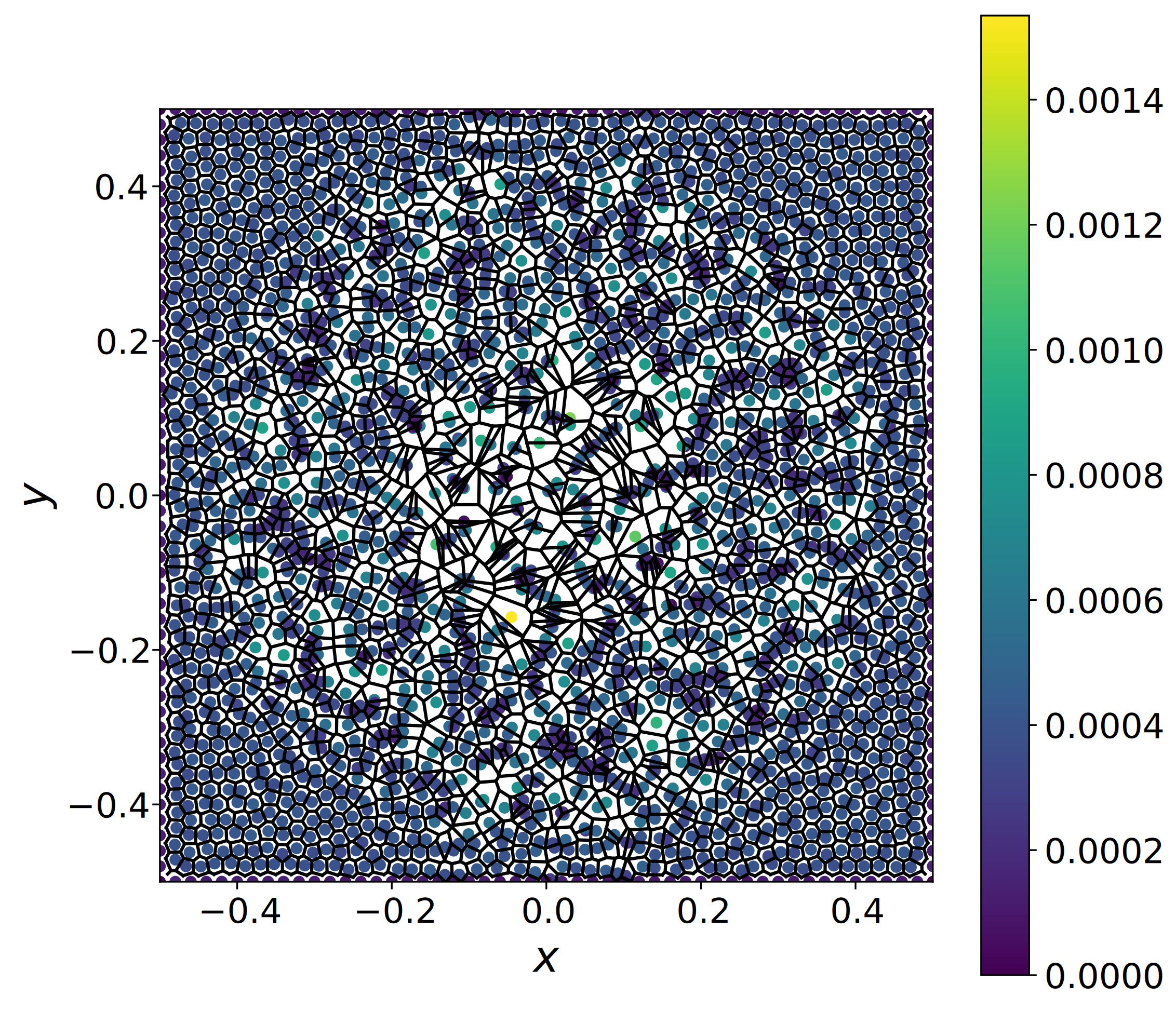} \\
	\includegraphics[width=0.75\linewidth]{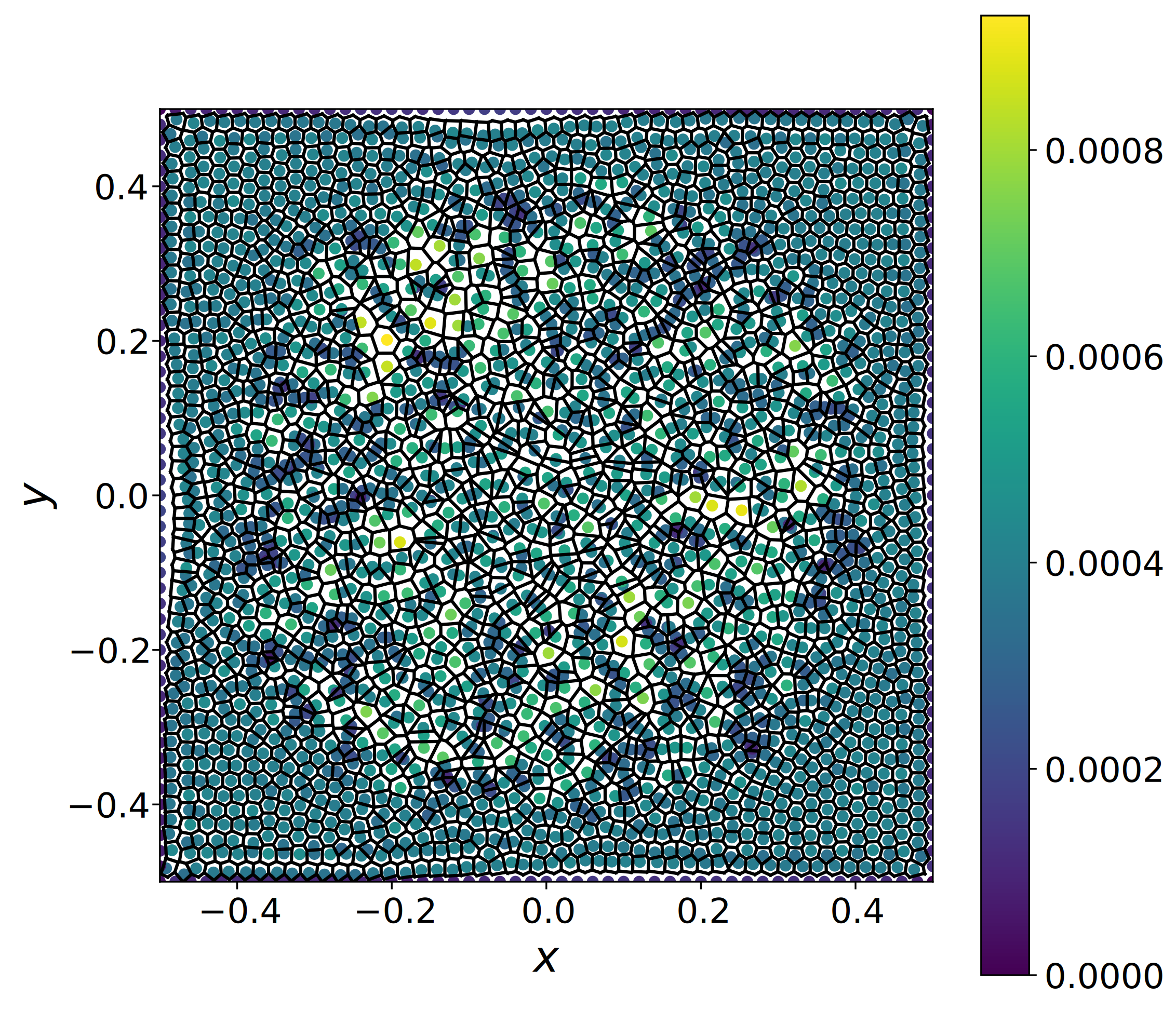}
	\caption{Snapshot of the system at $t=3 T$.
		Colors by volume.
		Methods shown:
		pFEM, and corrected pFEM.
	}
	\label{fig:snap_3.770}
\end{figure}

If we look at the last frame of the simulation, Fig.~\ref{fig:snap_3.770}, for only these two simulations, it seems clear that the additional forces have made a rather good job of keeping the particles more evenly distributed. Color is by volume in this Figure, also showing incompressibility has been better achieved with the corrected method. \added{Nevertheless, both methods present instabilities that are common in particle methods: pairing (more visible in the plain pFEM snapshot) and clumping into worm-like formations.}

\subsection{Taylor-Green vortex sheet}

\subsubsection{Setup}
\label{ss:simulations:setup_TG}

Another set of 2D simulations has been carried out for the Taylor-Green vortex sheet. We still consider an inviscid fluid, which will develop instabilities in the absence of viscosity, but these do not appear in the time range of our simulation, but at later times.

This is a periodic solution to the Euler equations that corresponds to a velocity field:
\begin{align*}
u_x =  &u_0  \sin (k x) \cos (k y) \\
u_y = -&u_0 \cos (k x)  \sin (k y)
\end{align*}
(the solution to the Navier-Stokes equation is similar, but involves a function that decays in time.) Here, $k=2\pi/L$ is the wave-vector for the spatial periodicity $L$. Every $L \times L$ square contains four vortices, with alternating rotations.

The corresponding pressure field is
\[
p = \frac{\rho u_0^2 }{4} \left( \cos (2kx) + \cos (2ky) \right) .
\]

An important practical difference with the Gresho vortex case is the boundary conditions. In order to implement periodicity, nine copies of the box are created when needed, in order to compute all the relevant geometrical information. This complements the CGAL libraries, which feature periodic Delaunay and Voronoi constructions but, at the moment, lack periodic constructions for power diagrams.

The simulations are otherwise similar the Gresho vortex case, with $49\times49= 2401$ particles on a square with $L=1$, particles initially placed on a square lattice. Results are given with no initial Lloyds procedure, since the problem seems to call for an initial square distribution (results with a Lloyds procedure have been checked to be nevertheless very similar.) The velocity and pressure scales are fixed by setting $u_0=1$. The rotation period, which is
$T = L / u_0$, is thus equal to $1$, and simulations are carried out for three complete periods. The time-step is again $\Delta t = 0.005$, for a Courant number Co=$0.25$. Results are discussed next for each method, with numerical details being identical to the Gresho vortex case, except where indicated.


\begin{figure}
	\centering
	\includegraphics[width=0.75\linewidth]{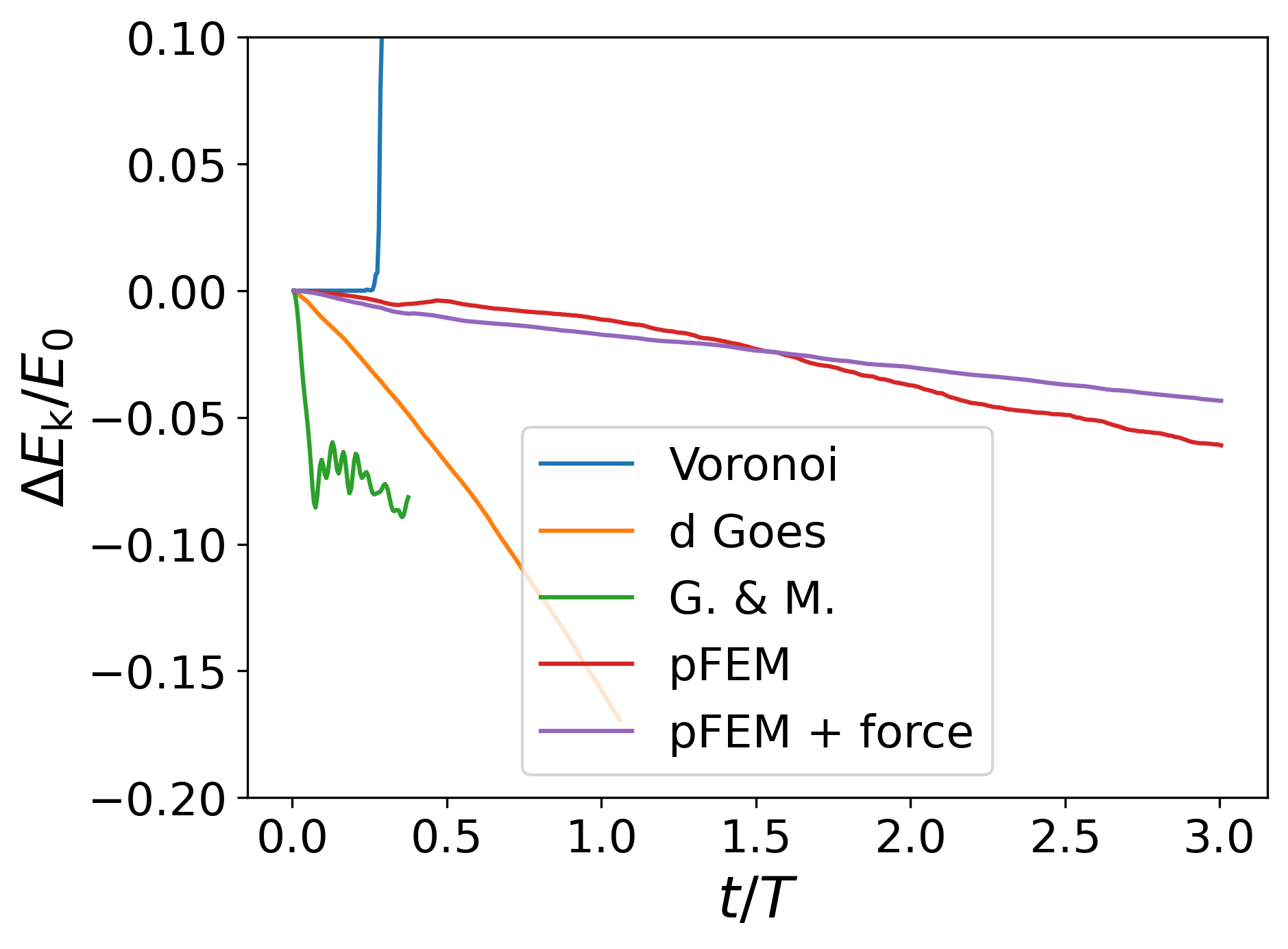} \\
	\includegraphics[width=0.75\linewidth]{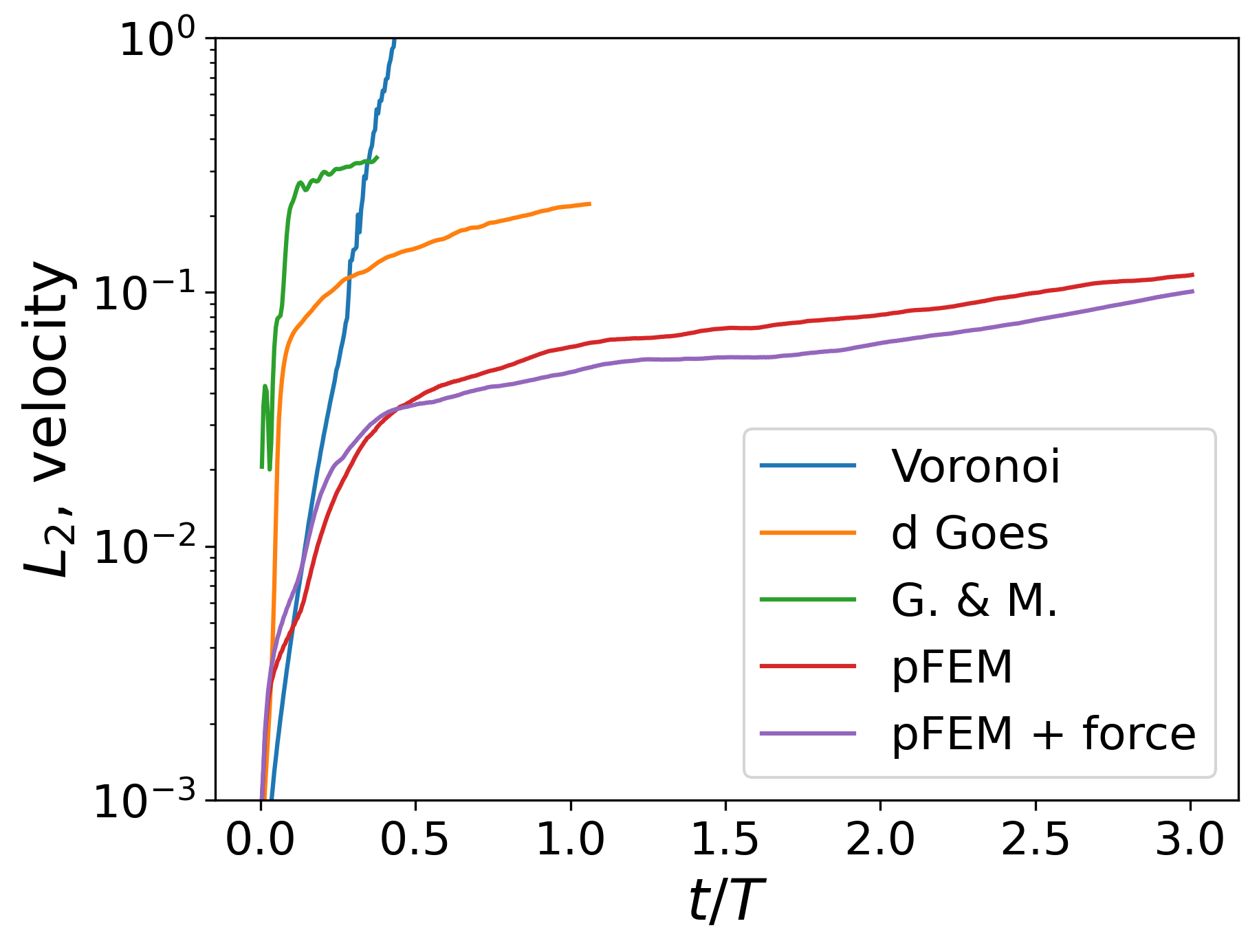} \\
	\includegraphics[width=0.75\linewidth]{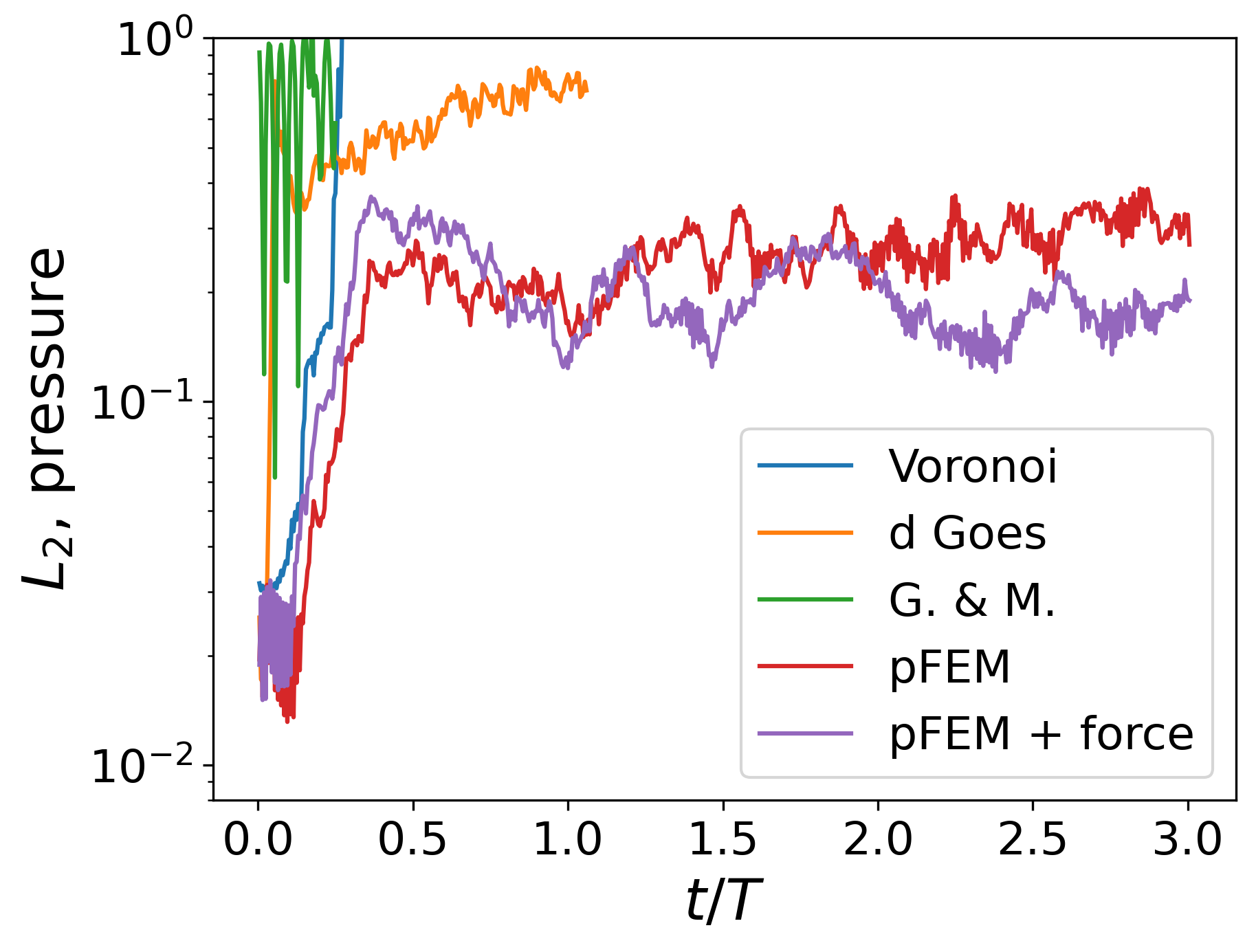}

	\caption{Taylor-Green vortex sheet simulations.
		Left: change in the relative value of the kinetic energy.
		Middle: relative $L_2$ distance between the velocity field and its theoretical value.
		Right: relative $L_2$ distance between the pressure field and its theoretical value.}
	\label{fig:ek_L2_TG}
\end{figure}

\subsubsection{Voronoi particles}

\begin{figure}
	\centering
	\includegraphics[	width=0.45\linewidth]{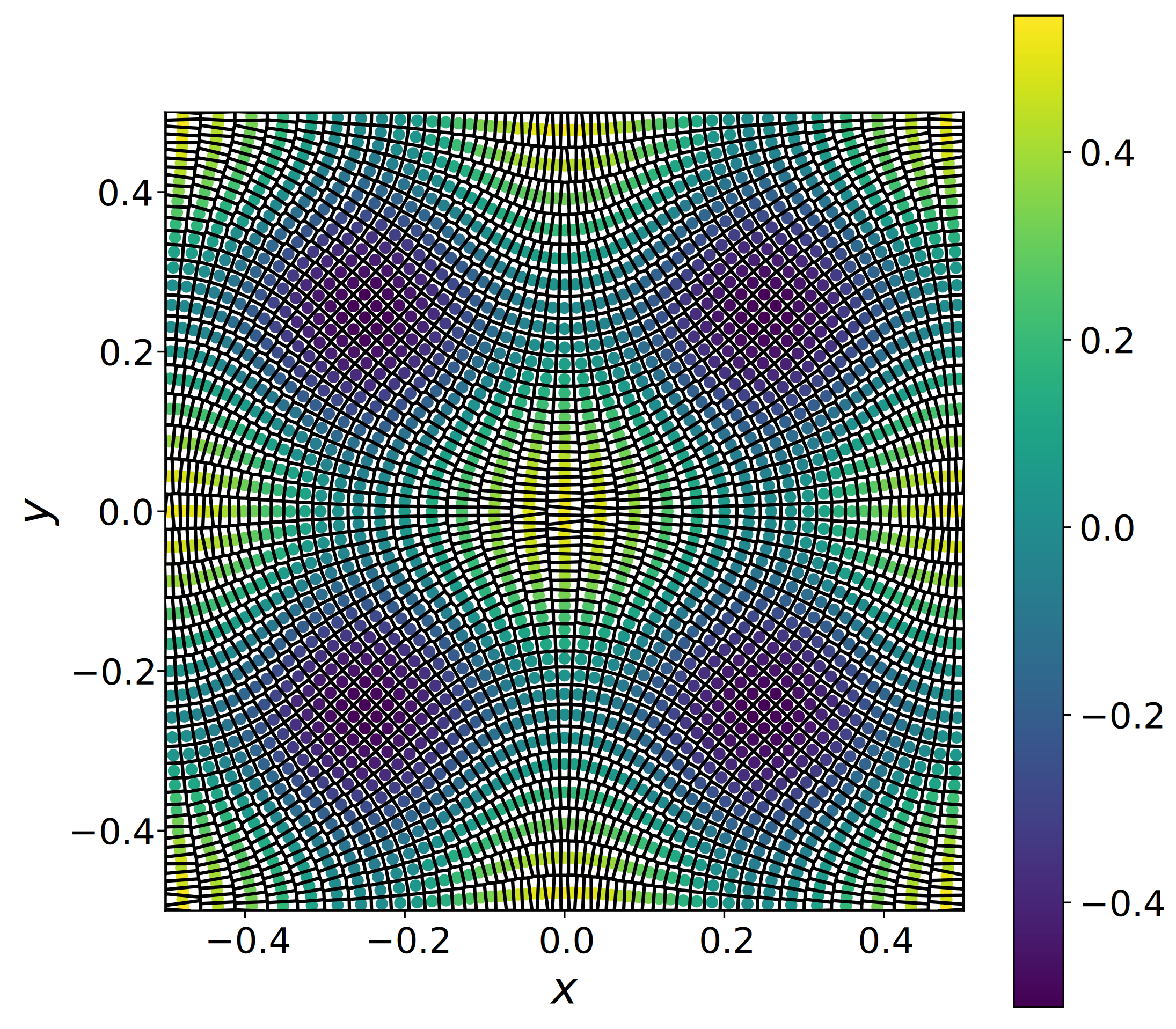} \hfil
	\includegraphics[	width=0.45\linewidth]{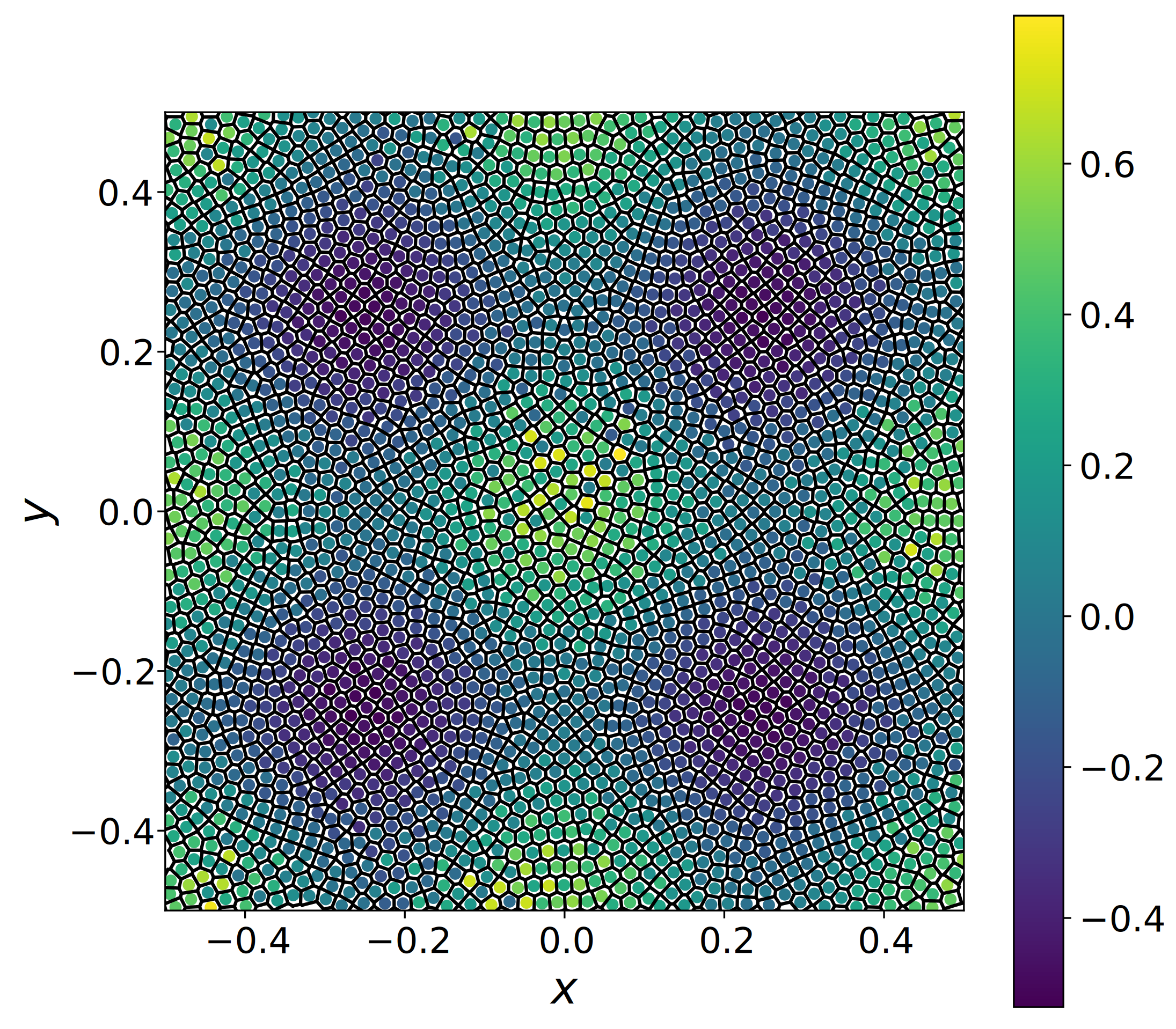} \\
	\includegraphics[	width=0.45\linewidth]{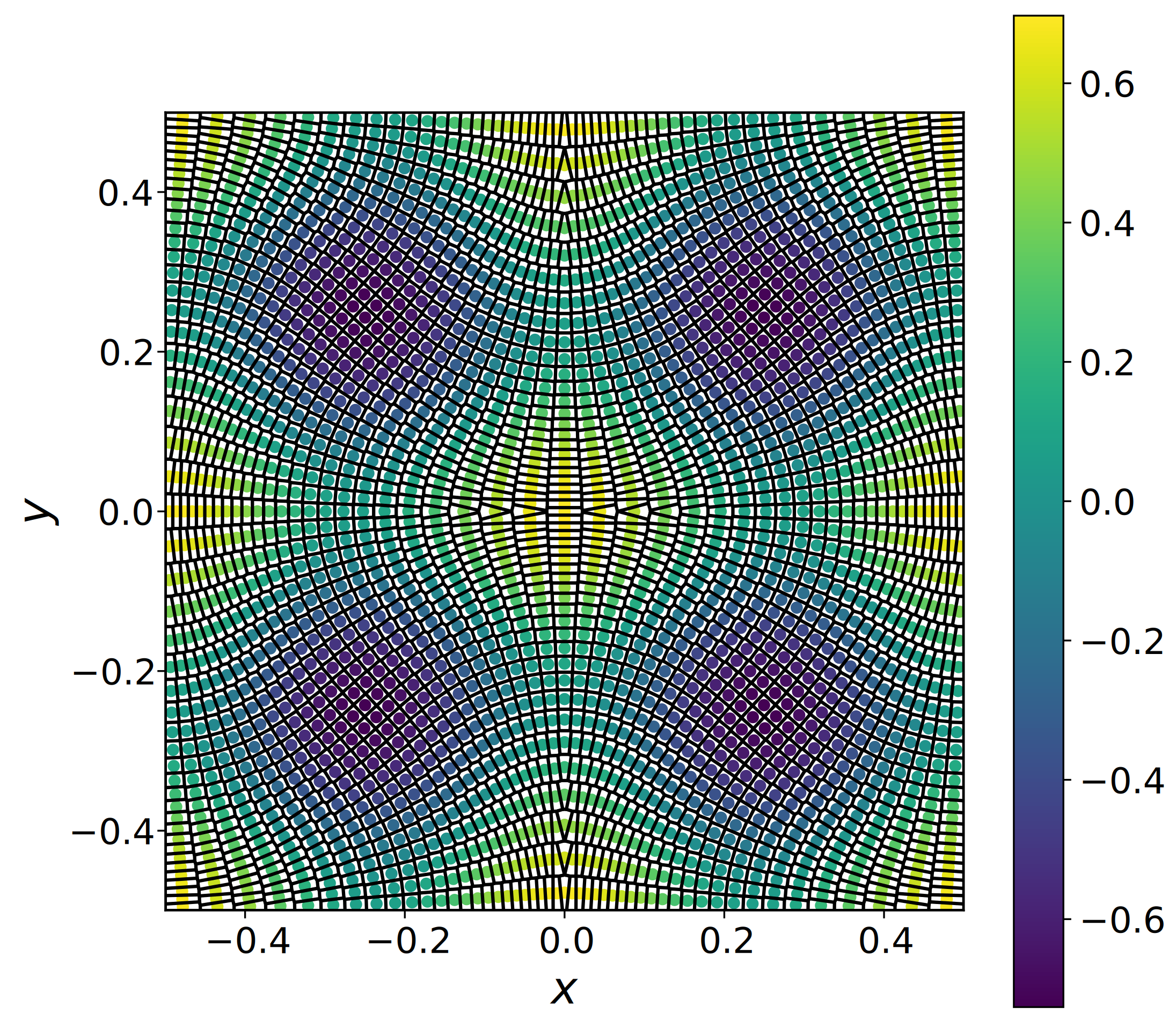} \hfil
	\includegraphics[	width=0.45\linewidth]{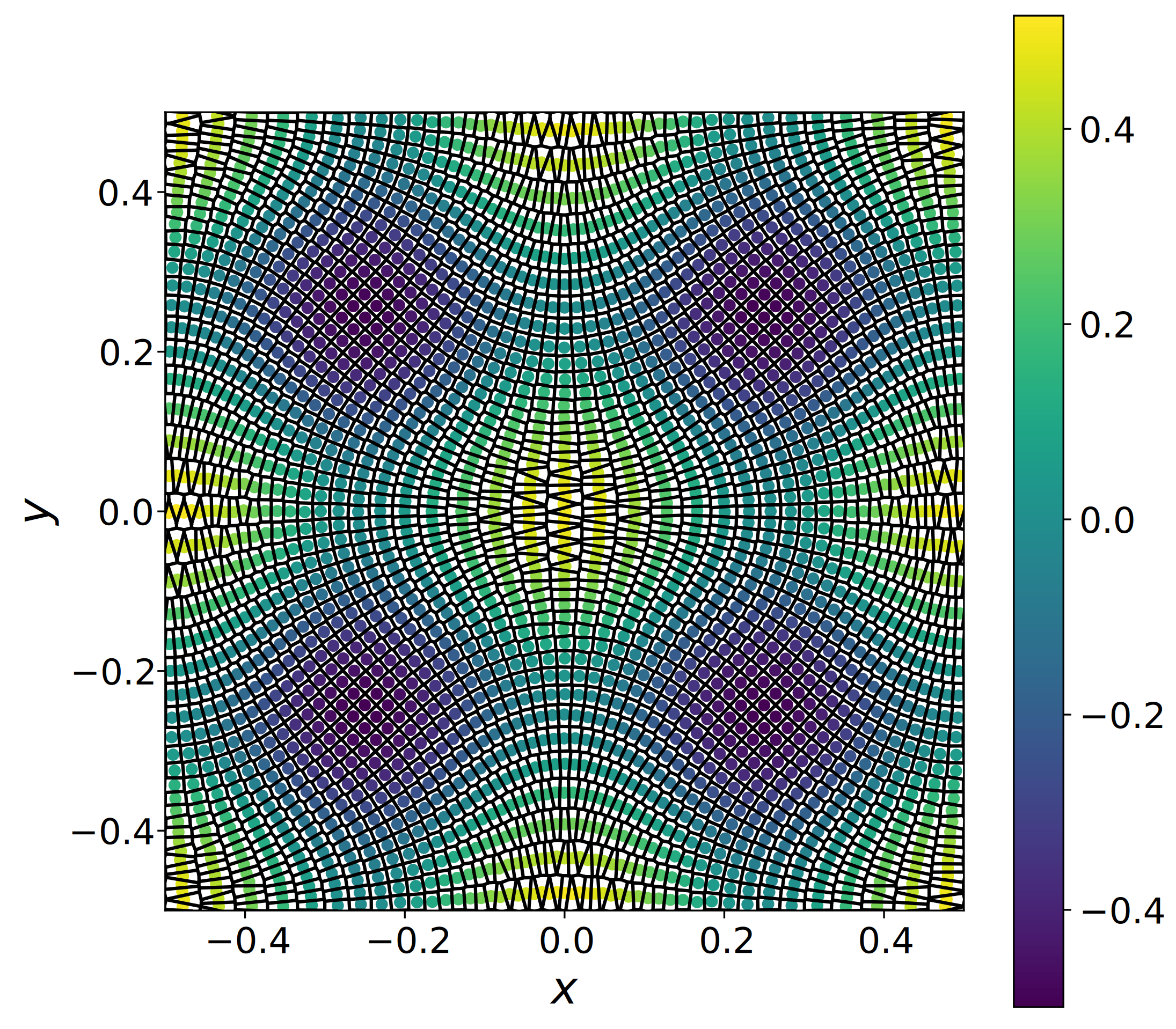} \\
	\includegraphics[	width=0.45\linewidth]{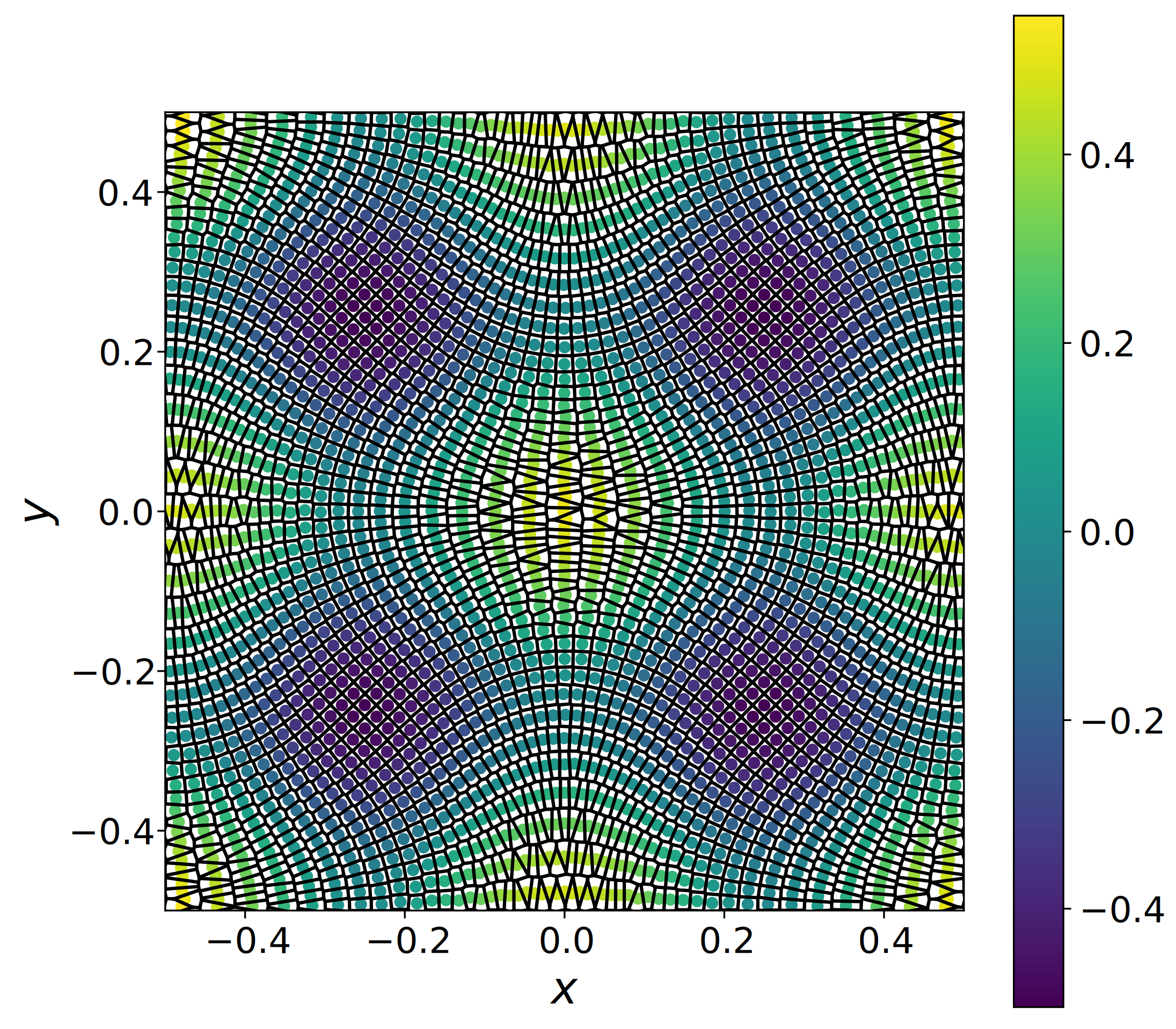}
	\caption{Snapshot of the Taylor-Green simulations at $t=T/8$.
		Colors by pressure.
		Methods shown:
		Voronoi particles, de Goes' method, Gallouët and Mérigot, pFEM, and corrected pFEM.
	}
	\label{fig:snap_0.125_TG}
\end{figure}

The VPD approach produces results that are similar to the Gresho vortex case.
In Fig. \ref{fig:snap_0.125_TG} snapshots of the system, at time $t = 0.125$, one-octave turn, are shown (some other simulations fail shortly after that time). The simulation is again acceptable in some ways, like the stable rotating cores.
Highly distorted cells are found at the zones where the fluid must squeeze between vortices, but this is expected. In fact, this case is a well-known test precisely because of this problematic zones.

For the sake of brevity, only averaged values are discussed for this case.
The kinetic energy in Fig. \ref{fig:ek_L2_TG} (left) shows good conservation
until a sudden blow-up. The simulation actually keeps running, but the velocity field show an unacceptable deviation (Fig. \ref{fig:ek_L2_TG} (middle)),
and so does the pressure field (Fig. \ref{fig:ek_L2_TG} (right)).

\subsubsection{Power particles}

For the de Goes power particle method, the particle arrangement of Fig. \ref{fig:snap_0.125_TG} seems reasonable. Notice the cells have undergone a shape change: moving the particles from the centroids no doubt implies a sort of moving Lloyds algorithm, by which the initial squares gradually become closer to hexagons.

In a manner similar to the Gresho vortex case, the kinetic energy is clearly dissipated, Fig. \ref{fig:ek_L2_TG} (left). The line is interrupted since the simulation fails at about $t=0.6$. This is due to the volume equalizing Newton-Raphson procedure (Eq.~\eqref{eq:NR}) failing to find the weights. This may be remedied by some more careful numerical technique, such as relaxation, but it indicates this particular case is difficult for an approach based on power particles. The velocity and pressure fields of Fig. \ref{fig:ek_L2_TG} (middle, right) are comparable to the Gresho vortex case.

\subsubsection{Gallouët and Mérigot}

Results are again very similar to to the Gresho vortex case, but the simulation
suffers from the same failure in the Newton-Raphson procedure as the de Goes algorithm. That is why results in Fig. \ref{fig:ek_L2_TG} are truncated, at an earlier time, actually, around $t=0.2$. The oscillations in energy and pressure that seem to be typical of this method are also seen in that figure (left, right).
Results shown correspond to a spring period $ \tau = 15 \Delta t $, a value that corresponds to $13\frac{1}{3}$ oscillations per cycle. Other values have been explored, but further from this value range stability is even worse.
In Fig. \ref{fig:snap_0.315}, the snapshot shows nodes cells that are closer in shape to the other methods, except the de Goes simulation.

\subsubsection{Standard, and corrected, pFEM}

At such early time, results from these two methods are hard to tell apart in Fig. \ref{fig:snap_0.125_TG}  (we remind that this time is forced by the failure of the GM method.)
Only at later times are results different. Fig. \ref{fig:ek_L2_TG} (left) shows little energy dissipation, with the corrected pFEM being superior at later times.
The velocity field, as seen in, Fig. \ref{fig:ek_L2_TG} (middle) is slightly better in the corrected formulation. It could be that the extra force included in the corrected method help a better numerical solution for the pressure, thus improving incompressibility. This could be confirmed by the pressure, which is seen Fig. \ref{fig:ek_L2_TG} (right) to remain within acceptable tolerances, and with lesser deviations in the corrected version.
A value of $\tau = 2\pi /\omega = 1.5 T $ was chosen for this case. Larger values, such as $\tau = 2 T $, yield worse results. This parameter therefore seems to be problem-dependent.

\section{Conclusions}
\label{ss:conclusions}

We have presented a unified framework in which the geometry and dynamics of a system of particles is treated simultaneously. The method is quite general, and we have used only its ``low-temperature'' limit, in which the geometry is forced to go to a low energy configuration, within the constraints imposed. If zeroth order consistency is required, this is the Voronoi diagram, while if first order is also required, it provides the Delaunay triangulation. If pressure is introduced, the Voronoi diagram leads to the method proposed by Gallouët and Mérigot (GM), based on power particles.
\replaced{Likewise, the Delaunay triangulation leads to the pFEM method. These two methods had been discussed in the literature, but the framework described here provides a robust pathway to deriving them. Moreover, it shows that an additional force may be present in pFEM.}{The Delaunay triangulation leads to the pFEM method, but now including an additional force.} 

These three methods (or, perhaps, two, since plain pFEM is corrected pFEM in the limit of a vanishingly small additional force), are the ones found within this framework. This means the other two we have discussed and simulated (Voronoi particles, and the de Goes procedure) would not be supported, and lie without it. Perhaps this explains their shortcomings: both may be unstable, and the de Goes method is dissipative. Both of are also close to GM, which is actually not too accurate either. It would seem that the zeroth order condition is too weak a requirement, since the first order is enforced, pFEM methods are derived, with results that are clearly superior.

Regarding additional adjustable parameters, VPD and dG have none, while GM and the corrected pFEM have a spring-like $\omega$ parameter. This offers flexibility, but of course a parameter-free approach is always preferable on fundamental grounds. In this regard, the fact that the volume equalizing procedure of Eq.~\eqref{eq:NR} looks so similar to a \replaced{standard Poisson pressure equation}{PPE} could provide a fixed value for the GM parameter. Unfortunately, our results in this direction have not been fruitful.

\added{%
While these two cases show that the additional force may improve the (already quite good) pFEM results, more particle methods (such as SPH) should be included in future benchmarks, and other cases should be considered. Those involving free surfaces are of particular importance, given the popularity of particle-based simulations for these situations. Its inclusion is, nevertheless, delicate --- for example, they have been only recently considered in the GM method \cite{levy_partial_2021}.}

We have also not targeted numerical cost. The Voronoi diagram and the dual Delaunay triangulation are evaluated at the same time in our codes. However, some schemes use explicit time integrators (dG, GM), while the others need an iterative procedure, which involves a number of evaluations per time-step. This number is usually only two or three, so the increased computational cost is likely acceptable, given the more accurate results. \added{From the point of view of computational cost, it is important whether other methods, such as SPH, may benefit from this perspective. For instance, procedures such as particle shifting, may be explained as a way to minimize some energy.}

\added{Conservation properties, of crucial importance, have been only briefly covered. Only VPD and plain pFEM can be expected to conserve energy --- it would seem that this feature is spoiled by the coupling with the geometry. It would be desirable to cast the problem in such a way that geometry requirements are realized as Lagrangian multipliers, so that their contribution can be null. Our attempts in this direction have been unfruitful, but that does not mean other approaches may be successful. The same applies, of course, to the conservation of linear momentum (which only plain pFEM can be shown to honor) and angular momentum. }

One final conclusion is that pFEM simulations may be improved and stabilized by the addition of a new force, which may be introduced at essentially zero cost. Whether this makes a significant impact on the simulation results (either physical, in an engineering setup, or visually appealing in a computer graphics environment) will need to be tested in future applications.

\section*{Methods}

Calculations are implemented in the C++ programming language, using intensive use of the CGAL libraries for computational geometry \cite{cgal:eb-22a}, especially their 2D triangulations \cite{cgal:y-t2-22a,cgal:py-tds2-22a}.
The codes used may be found in the open github repository \cite{github}.

\section*{Acknowledgments}
The author wants to express his gratitude for the financial support received by the Ministerio de Ciencia e Innovación (MCIU) under Grants RTI2018-096791-B-C21 and RTI2018-096791-B-C22:
 \textit{Hidrodinámica  de elementos de amor\-tigua\-miento del movimiento de aerogeneradores flotantes}.


\end{document}